\documentclass{article}

\usepackage{amssymb}
\usepackage{graphicx}
\newcommand{\be}{\begin{equation}}
\newcommand{\ee}{\end{equation}}
\newcommand{\bea}{\begin{eqnarray}}
\newcommand{\eea}{\end{eqnarray}}

\usepackage{color}
\usepackage{hyperref}

\begin{document}

\title{Theoretical Cosmology}

\author{A. A. Coley:\\
	Department of Mathematics and Statistics, 
	Dalhousie University,\\
	Halifax, Nova Scotia, B3H 4R2, Canada\\
	email: Alan.Coley@Dal.ca\\ \\
	G. F. R. Ellis:\\
	 Mathematics Department, 
	 University of Cape Town,\\
     Rondebosch, Cape Town 7701, South Africa\\
     email: george.ellis@uct.ac.za}

\maketitle
	
\begin{abstract}

We review current theoretical cosmology, including
fundamental and mathematical cosmology and physical cosmology (as well
as cosmology in the quantum realm),
with an emphasis on open questions.

\end{abstract}

\newpage
\tableofcontents

\newpage

\maketitle






\newpage
\section{Introduction}

Cosmology concerns the study of the large scale behavior of the Universe within a theory of gravity, 
which is usually assumed to be General Relativity (GR).
\footnote{Commonly used {\bf{Acronyms}} used in this paper include:
Cold dark matter (CDM).
Cosmic microwave background (CMB).
Einstein field equations (EFE).
Friedmann-Lema\'{i}tre-Robertson-Walker (FLRW).
Gravitational waves (GW).
General Relativity (GR).
Large scale structure (LSS).
Linear perturbation theory (LPT).
Loop quantum cosmology (LQC).
Loop
quantum gravity (LQG).
Primordial gravitational waves (PGW).
Primordial non-Gaussianities (PNG).
Quantum gravity (QG).} It has a unique nature that makes it a distinctive science in terms of its relation to both scientific explanation and testing.

\subsubsection{The uniqueness of the Universe}
There is only one Universe which we effectively see from one spacetime point (because it is so large) \cite{Ellis1971}. This is the foundational constraint in terms of both scientific theory (how do we distinguish laws from initial conditions?) and observational testing of our models:
\begin{itemize}
		\item We can only observe our Universe on one past light cone.
		\item We have to deduce four dimensional (4D) spacetime structure from a 2D image; 
distance estimations are therefore key.
		\item We can't see many copies of the universe to deduce laws governing how universes operate. Therefore, we have to compare the one universe with simulations of what might have been.
	\end{itemize}
\noindent	
This consequently leads to the important question
of what variations from our model need explaining and what are statistical anomalies that do not need any explanation (i.e., {\em{cosmic variance}}). This question arises, for example, regarding some  cosmic microwave background (CMB) anomalies.


\subsubsection{The background model} Cosmology is the study of the behaviour of the Universe when small-scale structures (such as, for example, stars and galaxies) can be neglected.
The ``Cosmological Principle'', which can be regarded as a generalization of the 
Copernican Principle,  is often assumed to be valid. 
This principle asserts that: 
{\em{On large scales the Universe can be well--modeled by a solution to Einstein's equations which is spatially homogeneous and isotropic.}} This implies that a 
preferred notion of cosmological time exists \footnote{Except in the degenerate cases of spacetimes of constant curvature (de Sitter, anti-de Sitter and Minkowski spacetimes). Such universe models do not correspond to the real Universe, which has preferred world lines everywhere 
\cite{Ellis1971}.} 
such that at each instant of time,  
space appears the same in all directions (isotropy) and at all places (spatial homogeneity)
on the largest scales.
This is, of course, certainly not true on smaller scales such as the astrophysical scales of galaxies,
and it would thus be better 
if the cosmological principle could be 
deduced  rather than assumed a priori 
(i.e., could late time spatial homogeneity and isotropy be derived as a dynamical consequence of the Einstein Field Equations (EFE) under suitable physical conditions and for appropriate initial data).
This has been addressed, in part, within the inflationary paradigm, when scalar fields are dynamically important in the early Universe.


The Cosmological Principle leads to a background Friedmann-Lema\'{i}tre-Robertson-Walker (FLRW) model, and the EFE determine its dynamics.
The concordance spatially homogeneous and isotropic FLRW model
(with a three-dimensional comoving spatial
section of constant curvature which is assumed simply connected) with a 
cosmological constant, $\Lambda$, representing the cosmological constant as an interpretation of
dark energy, and CDM is the acronym for cold dark matter (or  so-called $\Lambda$CDM cosmology
or {\em{standard cosmology}} for short), has been very successful in describing current observations.
Early universe inflation is often regarded as a
part of the concordance model. The background spatial curvature of the universe, often characterized by the normalized curvature parameter ($\Omega_{k}$), is predicted to be negligible by most of simple inflationary models.
Regardless of whether inflation is regarded as
part of the standard model, spatial curvature is often assumed zero.

\subsubsection{Inhomogeneous models} One of the greatest challenges in cosmology is understanding the origin of the structure of the Universe. 
An essential feature in structure formation is the study of inhomogeneities and anisotropies in cosmology. There are three approaches:
\begin{itemize}
\item Using exact solutions and properties where possible \cite{krameretal}. In particular, the Lema\^{i}tre-Tolman-Bondi (``LTB'')   spherically symmetric dust model has been widely used, while the ``Strenge Losungen'' approach of Ehlers, Kundt, Sachs, and Tr\"{u}mper at Hamburg provides a powerful method of examining generic properties of fluid solutions.  
\item Perturbed models where structure formation can be investigated, as pioneered by Lifschitz, Peebles, Sachs and Wolfe, Bardeen (see below).
\item Numerical simulations, mainly Newtonian, but now being extended to the GR by 
various groups.
\end{itemize} 
\noindent
In particular, this enables an investigation of 
the scalar-tensor ratio and CMB polarization, and
redshift space distortions and non-Gaussianities,
which are key to testing inflationary universe models. 

It is also important to consider the averaging, backreaction, and fitting problems relating the perturbed and background models. The main point here is that \textit{the same} spacetime domain can be modeled at different averaging scales to obtain, for example, models representing galactic scales $L_1$, galaxy cluster scales $L_2$, large scale structure scales $L_3$, and cosmological scales $L_4$, with corresponding metrics, Ricci tensors, and matter tensors; the issue then is, firstly, how the FE at different scales are related \cite{fit} and, secondly, how observations at these different scales are related \cite{Bertotti}.

\subsubsection{Perturbed models} In particular, 
the structure of the Universe can be investigated in cosmology via perturbed FLRW models. 
A technical issue that arises is the \textit{gauge issue}: how do we map the background model (smooth)  to a more realistic (lumpy) model? One must either handle gauge freedom by very carefully   delineating what freedom remains at each stage of coordinate specialisation, or use gauge covariant variables (see later).

Cosmic inflation provides a causal mechanism for the generation of primordial cosmological perturbations in our Universe,
through the generation of quantum fluctuations in the inflaton field which
act as seeds for the observed anisotropies in the CMB
and large scale structure (LSS) of our Universe.
Although inflationary cosmology is not the only game in town, it is the simplest and  perhaps the only scenario which is currently self-consistent from the point of view of
low energy effective field theory. The
recent Planck observations confirm that the primordial curvature
perturbations are almost scale-invariant and Gaussian. 
In the standard  cosmology, the primordial perturbations, corresponding to the seeds for the LSS, are chosen from a Gaussian distribution with random phases. This
assumption is justified based on experimental evidence, regardless of whether or not inflation is assumed.

Predictions arising for matter power spectra and CMB anisotropy power spectra can then be compared with observations; this is a central feature of cosmology today. Together with comparisons of element abundance observations with primordial nucleosynthesis predictions, this has turned cosmology from philosophy to a solid physical theory. 
Finally, quantum 
fluctuations of the metric during
inflation, imprinted in primordial B-mode perturbations of the CMB, are
perhaps the most vivid evidence conceivable for the reality of quantum gravity (QG) in the early history of our Universe.
Indeed, any direct detection of primordial gravitational waves (PGW) and
primordial non-Gaussianities (PNG) with the specific features predicted by
inflation would provide strong independent support to this framework.\\

In this article we review the philosophical, mathematical, theoretical, physical (and quantum)
challenges to the standard cosmology. For the most part, important non-theoretical issues (such as,
for example, experiments and data analysis) are not discussed.


\subsection{Fundamental issues}
Cosmology is a strange beast. On the one hand it has evolved into a mature science, complete with observations,
data analysis and numerical methods.
On the other hand it contains philosophical assumptions that are not always scientific; this includes, e.g., 
the assumption of spatial homogeneity and isotropy at large scales
outside our particle horizon and issues regarding  inflation and the  multiverse.
As well as philosophical questions, there are fundamental physical problems (e.g.,
what us the appropriate model for matter, and what is the applicability of coarse graining) 
as well as mathematical issues (e.g., the gauge invariance problem in cosmology). 
Indeed, many of the open problems in theoretical cosmology involve 
the nature of the origin and 
details of cosmic inflation.

\subsubsection{Open problems and GR}

Noted problems have always been of importance and part of the
culture in mathematics  \cite{OpenProb}.
The twenty-three problems by 
Hilbert \cite{Hilbert} are perhaps 
the most well known problems in mathematics. 
In addition, the set of fifteen problems presented in 
\cite{simon} nicely
illustrate a number of open
problems in mathematical physics.
The most important and interesting unsolved problems in fundamental theoretical physics include
foundational problems of quantum mechanics and
the unification of particles and forces and
the fine tuning problem in the 
quantum regime, the problem of quantum gravity and, of course,
the problem of ``cosmological mysteries''
\cite{gonitsora}.
However, it should be noted that some of them are in fact philsophical problems, in that they are not dealing with any conflict with observations.


We are primarily  interested here in problems which we shall refer to as problems in theoretical cosmology,
and particularly those that are susceptible to a rigorous treatment within mathematical cosmology.
Problems in GR have been discussed elsewhere  \cite{MathGR}. There are some problems in GR that are relevant
in cosmology, and theorems can be extended into the cosmological regime by including models with matter. 
For example, generic spacelike singularities, traditionally regarded as being cosmological singularities,
have been studied in detail \cite{SenovillaGarfinkle}. 
It is also of interest to extend mathematical stability results to the case of a non-zero cosmological constant
\cite{Dotti}.


\subsubsection{Philosophical issues}

Philosophical problems have always played an important role in cosmology \cite{Ellis2014}; e.g., are we situated at the center of the Universe or not and, even 
in the earliest days of Einstein, is the Universe static or evolving.
In addition, in cosmology the dynamical
laws governing the evolution of the universe, the classical EFE,
require boundary conditions to yield solutions. But in cosmology, by definition, there
is no “rest of the Universe” to pass their specification off to. The cosmological boundary
conditions must be one of the fundamental laws of physics.

There are a number of important philosophical issues that include the following:
There is only 
one Universe. Consistency of one model does not rule out alternative models.
What can a statistical analysis with only one data point tell us?
What is observable? Due to the existence of horizons, the Universe is only observed on or within our past light cone.
A typical question in cosmology is:
Why is the Universe so smooth. Must  a suitable explanation be in terms of `genericity'
(of possible initial conditions), or can specialness lead to a possible  explanation.
There is no physical law that is violated by fine tuning. Indeed, perhaps the Universe is fine-tuned due to anthropic reasons. However, there are many caveats in describing physical processes (e.g., inflation) in terms of naturalness. Indeed, in cosmology the whole concept of `naturalness' is suspect. Let us discuss some of these issues in a little more detail.


In {\em{observational cosmology}}, the amount of information that can be expected to be collected via astronomical observations is 
limited since we occupy a particular vantage point in the Universe; we are limited in what we can observe by visual and causal horizons (see discussion below). It can be argued
that the observational limit may be approached in the foreseeable future, at least
regarding some specific scientific hypotheses \cite{Ellis2014}. There is no certainty that the amount and types of information that can be collected will be
sufficient to test all reasonable postulated hypotheses statistically.
There is under-determination both in principle and in practice \cite{Butterfield, Ellis2014}.
This consequently leads to a natural view of model inference as inference to the best explanation/model, since some degree of explanatory ambiguity appears unavoidable in
principle; inference in cosmology is based on a {\em{Bayesian interpretation}} of probability
which includes a priori assumptions explicitly.


In {\em{physical cosmology}}, we are gravely compromised because we can only test physics directly up to the highest energies attainable by collisions at facilities such as the LHC, or from what we can deduce indirectly by cosmic ray observations. Hence we have to guess what extrapolation from known physics into the unknown is most likely to be correct; different extrapolations (e.g., string theory or loop quantum gravity) give different outcomes. As we cannot test directly the physics of inflation or of dark energy, theorists in fact rely mainly on Synge's g-method discussed below: we conclude matter has the properties we would like it to have, in order to fit with astronomical observations.

\subsubsection{Underlying theory}

It has been argued \cite{Sahlena}
that the measure problem, and hence model inference, is
ill defined due to ambiguity in the concepts of probability, in the situation where additional empirical observations cannot add any significant new information. However, inference in cosmological models
can be made conceptually well-defined  by extending the concept
of probability to general valuations (using principles of uniformity and consistency)  \cite{Sahlena}.

For example, an
important area is empirical tests of the inflationary paradigm which necessitates, in principle, the specification or derivation of an a priori probability of inflation occurring (``the measure problem'').
The weakness of all models of inflation is consequently in the initial conditions \cite{measure}.
To assert that the flatness of the Universe or the expected value for $\Lambda$ is predicted
by inflation is absolutely meaningless without such an appropriate measure 
(this is particularly true in the case of the multiverse \cite{SilkLimits}).

The fundamental problem is that the theory of inflation cannot be proven to be correct. 
Falsifying a ``bad theory'' (such as the the multiverse
“solution” to the cosmological constant problem \cite{Weinberg1987}) may be impossible \cite{SilkLimits}
{\footnote{The commonly accepted solution
to the mass hierarchy problem at the Planck scale necessitates an anti-de Sitter space-time and a negative  $\Lambda$. However, if  the sign of $\Lambda$ is allowed to have
anthropic freedom, the concept of using Bayesian constraints to yield a non-zero value for  $\Lambda$
from below must be discarded \cite{SilkLimits}.}}, since
parameters can be added without limit. 
But it should be possible to falsify a ``good theory'', like
inflation \cite{EllisSilk}. Perhaps the best way to make progress 
may be to probe the falsification of inflation, for which there is a robust predicted
CMB polarization signal
(induced by GW at the onset of inflation) \cite{SilkLimits}.

\subsubsection{Assumptions}

It is necessary  to make assumptions to derive models to be used for cosmological predictions and check with observational data. But what precisely are these assumptions and how do they
affect the results that come out;
e.g., is the
reason that small backreaction effects are obtained in computations 
because of the assumptions that are put in by hand at the beginning? We 
can only confirm the consistency of assumptions; we cannot rule out alternative explanations. 
The assumption of a  FLRW background (cosmological principle) on cosmological scales presents a number of problems.
There is no solid way to test spatial homogeneity, even in principle, by direct tests such as (redshift, distance observations), because we cannot control the possible time evolution of sources and so cannot be confident they are good standard candles (we do not, for example, have a solid understanding of supernova explosions and how they might depend on metallicity, or of radio source evolution). However, observations of structure growth on the one hand and matter-light interactions via the kinematic Sunyaev-Zeldovich effect on the other do indeed 
give rise to solid constraints on inhomogeneity \cite{Maartens_Homog,EllMaaMac}, and indicate that approximate spatial homogeneity does indeed hold within our past light cone.
Due to the existence of horizons, we can only observe the Universe on or within our past light cone (on cosmological scales).
Assumptions beyond the horizon
(Hubble scales) are impossible to test and so are, in effect, unscientific.

\subsubsection{Homogeneity scale}

The homogeneity scale is not actually theoretically determined, even in principle, in the
standard cosmological model. It is just ``pasted in" to the standard model a postieri to
help fit observations.
Even then, what is the derived homogeneity scale implied (from the statistics
observed). 
This question is important in the backreaction question.

There are a number of different approaches to the definition of a scale of statistical
homogeneity. Even if we consider the standard model setting, then the homogeneity
scale depends on the statistical measure used. But there are arguments 
that such a definition is not met and will never
be met  observationally \cite{sl09}.
Perhaps there is a different notion (e.g., using
ergodicity) of statistical homogeneity
in terms of an average positive density. But, any practical measure of
statistical homogeneity is not directly based on a fundemental relation, but rather
on the scale dependence of galaxy-galaxy correlation functions in observations \cite{dust}.

Observationally, and based on the 2-point correlation function,
the smallest scale at
which any measure of statistical homogeneity can emerge by the
current epoch is in
the range 70-120 $h^{-1} Mpc$.
Indeed, if all N-point
correlations of the galaxy distribution are considered, then the homogeneity
scale can only be reached, if at all, on scales above 700 $h^{-1} Mpc$ 
\cite{sl09} (also see \cite{DHB}).

\subsubsection{Local and global coordinates}

Perhaps, most importantly, 
what are the assumptions that underscore the use of an
inertial coordinate system  over a Hubble scale `background' patch
in which to do perturbation
computations or in specifying initial conditions for numerical GR evolution.
In particular, what are the assumptions necessary for the existence
of Gaussian normal coordinates and hence 
a `global' time and a `global' inertial (Cartesian and orthogonal) spatial  coordinate system
(and thus a $1+3$ split) on the `background' patch.
This necessitates an irrotational congruence of fluid-comoving observers and, of course, 
is related to a choice of lapse and shift and hence a well defined gauge.
And it essentially amounts to assuming that 
fluctuations propagate on a fixed absolute Newtonian background
(with post-Newtonian corrections).
Global inertial coordinates on a dynamical FLRW background in a GR framework  are not conceptually possible \cite{BC}.
A collection of spatially contiguous but causally disconnected regions which evolve according to GR on small scales do not generally evolve as a single collective background solution of GR on large cosmological scales. 
 
\subsubsection{Periodic boundary conditions in structure formation studies} 

In addition, what are the assumptions necessary for
periodic boundary conditions (appropriate on scales comparable to the
homogeneity scale) used in structure formation studies and numerical simulations? In particular,  periodic boundary conditions
impose a constraint on the global spatial curvature and
force it to vanish  \cite{Adamek,Macpherson}. Strictly speaking, the 
(average) spatial curvature is only zero 
in the standard cosmology in which the  FLRW universe
possesses the space-time structure
$R \times M^3$, in which $M^3$ is a three-dimensional spatial comoving simply connected infinite Euclidean 3-space of constant curvature. 
The EFE govern the local properties of space-time but
not the global geometry or the topology of the Universe at large.
Nonstandard models with a compact spatial topology (or {\em{small universes}}) which are periodic due
to topological identifications (and are hence not necessarily spatially flat)
are also of interest and have observational consequences \cite{Cornish}.
In particular, it has been shown that CMB data  are
compatible with the possibility that we live in a small Universe having the
shape of a flat 3-torus   with a sufficiently large volume  \cite{small}.

\subsubsection{Weak field approach} 

Finally,
what are the assumptions behind the weak field approach, the applicability of
perturbation theory (and use of Fourier analysis), Gaussian initial conditions, averaging and the neglect of backreaction? To different degrees they all assume a small (or zero)  spatial curvature.
In particular, all global averages of spatial curvature
are expected to coincide with that in the corresponding exact FLRW model to a high degree of accuracy
when averaging linear Gaussian perturbations.
In addition, in cosmology we can observe directions, redshifts, fluxes,  but not distances.
To infer a distance from observations in cosmology, we
always use a model. Hence the real space correlation function and its Fourier
transform, the power spectrum, are model dependent.

Essentially we conclude that within standard cosmology the spatial curvature is assumed to be zero (or at least very small and below the order of other approximations) in order for the analysis to be valid. In any case, the standard model cannot be used to {\em{predict}} a small spatial curvature.
We will revisit the issue of spatial curvature later.

\subsubsection{Quantum realm and multiverse}

Are there possible differences from GR at very small
scales that result from a theory of QG?
In particular, do they have any relevance in the cosmological realm,
and conversely what is the impact of cosmology on quantum mechanics
\cite{Hartle}.
For example,
are there any fundamental
particles that have yet to be observed and, if so, what are their properties?  
Do they (or the recently observed Higgs boson)
have any relevance for cosmology.  
There is also the issue of whether 
singularities can be resolved in GR by quantum effects and whether singularity theorems are possible in higher dimensions,
that are relevant in cosmology. 


Both QG and inflation motivate the idea of a multiverse, in which 
there exists a wide range of fundamental theories (or, at least, different versions of the same fundamental 
theory with different physical parameters) and our own Universe is but one possibility \cite{Carrmultiverse}.
In this scenario the question then arises as to why our own
particular Universe has such finely tuned properties that allow for the
existence of life. 
This has led to an explanation in terms of the so-called
anthropic principle,
which asserts that our Universe must have the properties it does because otherwise
we would not be here to ask such a question. 
The cosmology
of a multiverse leads to a number of philosophical questions. For example, is the multiverse even a scientific theory.



\subsection{Definition of a cosmological model}

A cosmological model has the following components \cite{Ellis1971}.

\noindent
{\em{Spacetime geometry:}}
The spacetime geometry
$({\bf M,g})$  is defined by a smooth Lorentzian  metric ${\bf g}$
(characterizing the macroscopic gravitational field) defined on a
smooth differentiable manifold ${\bf M}$ \cite{HawEll73}. The scale over which 
the cosmological model is valid should be specified.

\noindent
{\em{Field equations and equations of motion:}}
To  complete the definition of  a cosmological model,
	we must specify
	the physical relationship (interaction) between the macroscopic
	geometry and the matter fields, including how the matter responds
	to the macroscopic geometry.
	We also need to know the
	trajectories along which the cosmological matter and light moves.
In standard theory, the space-time metric, ${\bf g}$, is determined by the 
matter present via the EFE: 
	\begin{equation}\label{eq:EFE}
	G_{ab} := R_{ab} - \frac{1}{2}R g_{ab} = \kappa T_{ab} -\Lambda g_{ab}
	\end{equation}
where the total energy momentum tensor, $T_{ab}$, is the sum of the stress tensors of all matter components present: $T_{ab} = \sum_{(i)}T_{ab}^{(i)}$, $\kappa$ is essentially the gravitational constant, and $\Lambda$ is the cosmological constant. In colloquial terms: \textit{Matter curves spacetime}. Because of the Bianchi identities, $R_{ab[cd;e]}=0$,  the definition on the left of (\ref{eq:EFE}) implies the identity $G^{ab}_{\,\,\,\,;b} = 0$ and hence, provided $\Lambda$ is indeed constant, that: 
\begin{equation}\label{eq:cons}
G^{ab}_{\,\,\,\,;b} = 0 \,\,\, \Rightarrow T^{ab}_{\,\,\,\,;b} = 0;
\end{equation}
that is, energy-momentum conservation follows identically from the FE (\ref{eq:EFE}). The covariant derivatives in (\ref{eq:cons}) depend on the space-time geometry, so in colloquial terms: \textit{Space-time tells matter how to move.} The key non-linearity of GR follows from the combination of these two statements, and the fact that $R_{ab}$ is a highly non-linear function of $g_{ab}(x^i)$.

In GR a test particle follows a timelike or null geodesic. But a system that behaves as point
particles on small scales may {\em not necessarily} do so on larger scales. 
That is, if the particles traverse timelike geodesics in the
microgeometry, in principle, the macroscopic (averaged) matter
need not follow timelike geodesics of the macrogeometry. 
However, the fundamental congruence is, in essence, the average of the
timelike congruences along which particles move in the
microgeometry, and defining the effective conserved 
energy-momentum tensor ${T}^{a}_{~b}$ through the EFE 
ensures timelike geodesic motion.
In addition, the
(average) motion of a photon is not necessarily on a null geodesic in the
averaged macrogeometry, which will affect observations.

\noindent
{\em{Matter:}}
We require a
consistent model for the matter on the characteristic cosmological
(e.g., averaging) scale, and its appropriate (averaged) physical
properties. Differentiation between the
gravitational field and the matter fields is known not to be scale invariant and, in particular,  
a perfect fluid is not a scale invariant phenomenon \cite{LL};
averaging in the ``mean field theory" in the presence of  gravity changes the
equation of state of the matter \cite{Deb}. 
In this framework all of the qualitative  effects of averaging are absorbed
into the redefined  effective energy-momentum tensor ${T}^{a}_{~b}$ and
the redefined effective equation state of the macro-matter, 
where $T^{a}_{~b}$ is conserved relative to the
macrogeometery. The definition of the Landau frame for any combination of matter fields and radiation is invariant when matter and matter-radiation interactions take place due to local momentum conservation.

\noindent
{\em{Timelike congruence:}}
There is a preferred unit timelike
congruence ${\bf u}$ ($u^a u_a = -1$),
defined locally at each event, 
associated with a family of
fundamental observers (at late times) or the average motion of energy (at earlier times).
In the case that there is more than one
matter component, implying the existence of more than one fundamental macroscopic timelike
congruence, we can always identify a fundamental macroscopic timelike
congruence represented by the 4-velocity of the averaged
matter in the model; i.e., the matter fields admit a formulation in
terms of an averaged matter content which defines an average
(macroscopic) timelike congruence.
This then leads to a covariant $1+3$ split of spacetime \cite{Ellis1971}.
Mathematically this implies that the spacetime is topologically restricted and is
\emph{$\mathcal{I}$-non-degenerate}, and consequently the spacetime is
uniquely characterized by its scalar curvature invariants \cite{CHP}.
For example, for a perfect fluid  ${\bf u}$ is the timelike eigenfunction of the Ricci tensor. 

Observationally, this cosmological rest frame is determined as the frame wherein the CBR dipole is eliminated (the Solar System is moving at about $370 km/sec$ relative to this rest frame). Note that the existence of this preferred rest frame is an important case of a \textit{broken symmetry}: while the underlying theory is Lorentz invariant, it's cosmologically relevant solutions are not (in particular, at no point in the history of the universe is it actually de-Sitter -- with its 10-dimensional symmetry group -- much less anti-de Sitter).

\noindent
{\em{A note on modified theories of gravity:}} Let us make a brief comment here. A key issue is whether GR is, in fact, the correct theory of gravity, especially on galactic and cosmological scales. Recent developments in testing GR on cosmological scales 
within modified theories of gravity were reviewed in \cite{Ishak,Clifton_Ferreira}.
In particular, modified gravity theories have played an important role in the dark energy problem.
Many questions can be posed in the context of modified gravity theories which include, 
for example, the general applicability of the BKL behaviour in the neighborhood of a cosmological 
singularity. We will not discuss
such questions here, except for the particular question of whether
isotropic singularities are typical in modified gravity theories.

\subsection{Problems in mathematical cosmology}

In GR, a sufficiently differentiable
4-dimensional Lorentzian manifold is assumed \cite{HawEll73}.
The Lorentz metric, g, which characterizes the causal structure of M, is
assumed to obey the EFE, which constitute a 
hyperbolic system of  quasi-linear partial differential
equations which are coupled to additional partial differential equations describing
the matter content \cite{Rendall2002}.
The Cauchy problem is of particular interest, in which
the unknown variables in the constraint equations of the governing EFE,  consisting of a
three-dimensional Riemannian metric
and a symmetric tensor (in addition to initial data for any matter fields
present), constitute the initial data for the remaining EFEs.
Primarily the vacuum case is considered in attempting to prove theorems in GR, but this is not the case of relevance in cosmology. Viable cosmological models contain both matter and radiation, which in physically realistic cases then define a geometrically preferred  timelike 4-velocity field
\cite{Ellis1971} which,  because of (\ref{eq:EFE}), is related to an eigenvector of the matter stress tensor (which is unique if we assume realistic energy conditions \cite{HawEll73}).


The EFE
are invariant under an arbitrary change of coordinates
(general covariance), which
complicates the way they should be formulated in order 
for global properties to be investigated \cite{LARS99}.
The vacuum EFEs are not hyperbolic in the normal sense. But
utilizing general covariance,  in harmonic coordinates
the vacuum EFEs do represent a quasi–linear hyperbolic
system and thus the Cauchy problem is indeed well posed and local existence is guaranteed 
by standard results \cite{CB69}. 
It can also be shown that if the constraints (and any
gauge conditions) are satisfied initially, they are preserved by the evolution.
Many analogues of the results in the vacuum case
are known for the EFE coupled to different kinds of matter, including perfect fluids, 
gases governed by kinetic theory, scalar
fields, Maxwell fields, Yang-Mills fields, and various combinations of these. 
Any results obtained for (perfect) fluids are generally
only  applicable in  restricted circumstances such as, for example, when the energy density is uniformly bounded away from zero (in the region of interest) \cite{Rendall2002}.
The existence of global solutions for models with more exotic matter, such as stringy matter,
has also been studied \cite{Narita}.

\subsubsection{Singularity theorems}

The concepts of geodesic incompleteness
(to characterize singularities) and closed trapped surfaces \cite{Penrose1979}
were first introduced in the
singularity theorem due to Penrose \cite{Penrose65}.
Hawking and Ellis \cite{HawEll68} then proved that closed trapped surfaces will indeed exist in the reversed direction of time in cosmology, due to the gravitational effect of the CMB.
Hawking subsequently realized that closed trapped
surfaces will also  be present in any expanding Universe  in its past, which would then inevitability 
lead to an initial singularity 
under reasonable conditions   within GR
\cite{Hawking1966}. 
This led to the famous Hawking and
Penrose singularity theorem \cite{PenroseHawking}.

The singularity theorems prove the inevitability of spacetime
singularities in GR under rather general conditions
\cite{Penrose65,PenroseHawking}, but they say very little about the actual nature of generic
singularities. We should note that there are generic spacetimes which do not have singularities
\cite{Senovilla2012}.
In particular, the proof of the Penrose singularity theorem does not
guarantee that a trapped surface will occur in the evolution. 
It was proven  \cite{Christodoulou2009} that for
vacuum spacetimes a trapped surface can, indeed, form dynamically from regular
initial data free of any trapped surfaces. 
This result was subsequently generalized in
\cite{Klainerman2014,Klainerman2012}.  A number of
questions still exist, which include proving more general
singularity theorems with weaker  energy conditions and with weaker differentiability, and 
determining any relationship
between geodesic incompleteness and the divergence of
curvature \cite{Senovilla2012}.
But perhaps the most important open problem within GR is cosmic censorship \cite{MathGR}.

\subsubsection{Bouncing models}

Using exotic matter, or alternative modified theories of gravity,
can classically lead to the initial cosmological (or “big bang”) singularity being replaced
by a  “big” {\em{bounce}}, a smooth transition from
contraction to  an expanding universe \cite{Brandenberger},
which may help to resolve some fundamental problems in cosmology. 
Bounce models have utilized ideas like branes and extra dimensions 
\cite{Khoury}, Penrose's  conformal cyclic cosmology \cite{Penrosebooks} (which leads to an interest in an isotropic singularity),    
string gas \cite{BrandenbergerVafa}, and others 
\cite{Brandenberger,Bruni_bounce}.

The matter bounce scenario faces significant problems. In particular, the contracting phase is unstable against
anisotropies \cite{Cai} and inhomogeneities \cite{Penrosebooks1}. In addition, there is no suppression of GW compared to cosmological perturbations,
and hence the amplitude of GW (as well as possible induced 
non-Gaussianities) may be in excess of the observational bounds. 
In a computational  study of
the evolution of adiabatic perturbations in a nonsingular bounce  within the ekpyrotic cosmological
scenario \cite{EKp}, 
it was  shown that the bounce is disrupted in regions 
with significant spatial inhomogeneity and anisotropy compared with the background
energy density, but is achieved in regions that are relatively spatially homogeneous and
isotropic.

The specially fine-tuned and simple examples studied to date,
particularly those based on three spatial dimensions, scalar fields and, most importantly, a non-singular bounce
that occurs at densities well below the Planck scale where QG effects are small
\cite{Ijjas2}, are arguably
instructive in pointing to more physical
bouncing cosmological models, and may present realistic
alternatives to inflation to obtain successful structure formation
(which we will discuss below).

The precise properties of a cosmic bounce depend upon the way in which it is
generated, and many mechanisms have been proposed for this both classically and non-classically. 
Bounces can occur due to 
QG effects associated with string theory \cite{Turok} and loop
quantum gravity  \cite{Bojowald,Ashtekar}.
In particular, in loop quantum cosmology there is
a bounce when the energy density reaches a maximum value of
approximately one half of the Planck density
(although  it is also possible that bounces occur without a QG regime ever occurring \cite{bounce}, because if inflation occurs, the inflaton field violates the energy conditions needed for the classical singularity theorems to be applicable). We will discuss this in more detail later.


\subsubsection{Mathematical results} 

Some applications in GR can be studied via Einstein-Yang-Mills (EYM) theory
(which is relevant to cosmological models containing Maxwell fields and form fields 
and is perhaps a prototype to studying fields in, for example, string theory). Mathematical results when 
generalized to Maxwell and YM matter in 4D \cite{Olinyk} are known 
(and have been studied in two  dimensions less
by {\em{wave maps}} with values on spheres \cite{Bizonwavemaps,AnderssonG}).

Global existence in Minkowski spacetime, assuming initial data of
sufficiently high differentiability, 
was first investigated in \cite{differentiability}.
The uniqueness theorem for the 4D  Schwarzschild spacetime was presented in \cite{Bunting}.
The uniqueness theorem for the Kerr spacetime was proven in \cite{Carter}.
In the non-vacuum case  the uniqueness of the rotating electrically charged black hole solution
of Kerr-Newman has not yet been generally proven  \cite{Newman65}.
Once uniqueness has been established, the
next step is to prove stability under perturbations.
Minkowski spacetime has been shown to be globally stable \cite{ChristodoulouKlainerman90,Christodoulou93}.

\subsubsection{Extension to cosmology}

Many of these problems in GR can be extended to the cosmological realm \cite{MathGR}.
The uniqueness and stability of solutions to  the  
EFE in GR are important, 
\footnote{
A full proof of  the  linear
stability of Schwarzschild spacetime has recently been established \cite{DHR}.
The non-linear stability of  the  Schwarzschild spacetime
is still elusive \cite{Heusler} (however, see \cite{KlainermanSzeftel}). 
Proving  the  non-linear stability of Kerr 
has become one of  the  primary areas of
mathematical work in GR \cite{Christodoulou93,Shlapentokh}).
All numerical results, and current observational
data, provide evidence
that  the  Kerr (and Kerr-Newman) black holes are non-linearly stable
\cite{Zilho}.}
and can be generalized to 
cosmological spacetimes (with a cosmological constant). 
Generic spacelike singularities are traditionally referred to as being cosmological singularities \cite{SenovillaGarfinkle}.
In particular, 
the stability of de Sitter spacetime will be discussed later.
There are also a number of  questions  in the quantum realm \cite{OpenProb}, such as
singularity resolution in GR by quantum effects and  higher dimensional models,
which are of interest in cosmology.

In essence the perturbation studies leading to theories of structure formation are stability studies of FLRW models.  With ordinary equations of state, initial instabilities will grow but with a rate that depends on the background model expansion. Thus if there is no expansion, inhomogeneity will grow exponentially with time; with power law expansion, they will grow as a fractional power of time; and with exponential expansion, they will tend to freeze out. However, the way this happens depends on the comoving wavelength of the perturbation relative to the scale set by the Hubble expansion rate at that time.\footnote{{Often erroneously called the `Horizon'. It has nothing to do with causality, i.e. with effects related to the speed of light.}} 
These studies hold while the perturbation is linear, and have been heroically extended to the non-linear case (see later). However numerical simulations are required for the strongly non-linear case \cite{Adamek}.

\subsubsection{Computational cosmology}

Numerical calculations have always played a central role in GR. Indeed, 
numerical computations support many of the conjectures in GR and their counterparts in cosmology and have led to a number of very important  theoretical
advances \cite{MathGR}. 
For example, the investigation of the mathematical stability of AdS spacetime includes fundamental 
numerical work and
cosmic censorship is supported by numerical computations.
In addition, the role of numerics in 
the understanding of  the BKL dynamics, 
and in various other problems in cosmology and higher dimensional gravity, has been important.
In fact, numerical computations are now commonly used to
address fundamental issues within full GR cosmology\cite{Computational,Bentivegna,Giblin,Adamek,Macpherson,Adamek18}.

\subsection{Cosmological observations}

What turns cosmology from a mathematical endeavour to a scientific theory is its ability to produce observational predictions that can be tested. Since the initiation of cosmology as a science by Lema\"{i}tre in 1927 \cite{Lemait27}, telescopes of ever increasing power, covering all wavelengths and both Earth-based and in satellites, have led to a plethora of detailed tests of the models leading to the era of ``precision cosmology''. The tests are essentially of two kinds: direct tests of the background models based on some kind of ``standard candle'' or ``standard ruler'', and indirect tests based on studying the statistics both of structures (inhomogeneities) on the one hand, and their effects on the CMB on the other. Both kinds of results produce broadly concordant results, but the latter give tighter restrictions on the background model than the former, because what kinds of structures can form depends on the dynamics of the background model.

\noindent
{\em{The basic restricton:}} The basic observational restriction in cosmology is that given the scales involved, we can only observe the Universe from one space-time event (``here and now'') \cite{Ellis1971}. This would not be the case if the Universe were say the size of the Solar System, but that is not the case: a key discovery has been the immense size of the Universe, dwarfing the scales of galaxies which themselves dwarf the scale of the Solar System. This leads to major limits on what is observable, because of \textit{visual horizons} for each kind of radiation or particle: for example, the CMB is observed on single surface (two-sphere) of last scattering. The furthest matter we can observe can be influenced by matter even further out, but such indirect effects are limited by the \textit{particle horizon}: the furthest matter that can have had causal influence on us by influences travelling to us at speeds limited by the speed of light since the start of the Universe.

\subsubsection{Anomalies} 

Within theoretical cosmology there needs to be an adequate explanation of observational anomalies, which are bound to occur as we make ever more detailed models of the structures and their effects on the CMB. 
Geometric optics must be utilized and model independent observations are sought.
In general, data
analysis and statistical methods are not discussed here. However, observations do, of course, lead to theoretical questions. Are there important neglected selection/detection effects  \cite{Disney}; i.e., what else can exist that we have not yet seen or detected?
Observations sometimes lead to ridiculous predictions (e.g., $w < -1$; phantom matter); care must be taken not to
be led into unphysical parameter space. Appropriate  explanations of observational anomalies
may well lead to new fundamental  physics and questions.

The standard cosmology
has been extremely successful in describing current observations,
up to various possible
anomalies and tensions \cite{tension}, and particularly some statistical features in the CMB \cite{planck2018} and the existence of structures on gigaparsec
scales such as the cold spot and some super-voids \cite{Finelli}.

Although primordial
nucleosynthesis has been very successful in accounting for the abundances of
helium and deuterium, 
lithium has been found to be overpredicted by a factor of about three \cite{lithum}. Lithium, along with deuterium,
is destroyed in stars, and consequently it's observation constitutes evidence (and a measure) of the primordial abundance after
any appropriate
corrections. 
To date there has been some claims of relief in this tension, but there is no satisfactory resolution
of the lithium problem.

A seldom asked question is whether the CMB and matter dipoles are in agreement \cite{EllisBaldwin}.
Tests of differential cosmic expansion on such scales rely
on very large  distance
and redshift catalogues, which are noisy and are subject to numerous
observational biases which must be accounted for. 
In addition, ideally 
any test should be performed in a model independent
manner, which requires removing the FLRW assumptions
that are often taken for granted in
many investigations. 
To date, such a model independent test has
been performed for full sky spherical averages of local expansion \cite{WiltshireHubble}, using
the COMPOSITE 
and Cosmicflows-II  catalogues; it was
found with very strong Bayesian evidence that the spherically
averaged expansion is significantly more uniform in the rest
frame of the Local Group (LG) of galaxies than in the standard
CMB rest frame.
It was subsequently shown by
that this result is consistent with Newtonian N-body simulations
in the standard cosmology framework \cite{Kraljic}. The future of such tests is discussed in \cite{Maartens_dipole}, concluding that the amplitude of the matter dipole can be significantly larger than that of the CMB dipole.
Its redshift dependence encodes information on the evolution of the Universe and on the tracers.

Perhaps more controversially, it has also been suggested that a
``dark flow'' may be responsible
for part of the motion of large objects that has been observed.
An analysis of the local bulk flow of
galaxies indicates a lack of 
convergence to the CMB frame
beyond 100 Mpc \cite{Kashlinsky}, which contradicts standard cosmological
expectations. Indeed, 
there is an anomalously high and approximately constant bulk flow of roughly 250 km/s extending
all the way out to the Shapley supercluster at approximately 260 Mpc, as indicated by
low redshift supernova data.
Furthermore, there is a 
discrepancy which has been confirmed by
6dF galaxy redshift data
\cite{Sarkar}.

\subsubsection{Tension in the Hubble constant}

The recent determination of the local value of the Hubble constant based on direct measurements of supernovae made with the Hubble Space Telescope \cite{R16} is now $3.3$ standard deviations higher than the value derived from the most recent 
data on the power spectrum temperature
features in the CMB provided by the Planck satellite in a $\Lambda$CDM model. 
Although it is unlikely that there are no systematic errors (since the value of the Hubble constant has historically been a source of controversy), the difference might be a pointer towards
new physics \cite{tensionRiess}. So this is perhaps the most important anomaly that needs to be addressed.

Although a large number of authors have proposed several different mechanisms to explain this tension, after three years of improved analyses and data sets, the tension in the Hubble constant between the various cosmological datasets not only persists but is even more statistically significant. The recent analysis of \cite{R16} found no compelling arguments to question the validity of the dataset used.
Indeed, the recent determination of the local value of the Hubble constant by Riess {\em{et al.}} in 2016 \cite{R16} of $H_0=73.24 \pm 1.74 km s^{-1} Mpc^{-1}$  at $68 \%$ confidence level  is now about $3$ standard deviations higher than the (global) value derived from the earlier 2015 CMB anisotropy data provided by the Planck satellite assuming a $\Lambda$CDM model \cite{planck2015}. This tension only gets worse when we compare the Riess {\em{et al.}} 2018  value of $H_0=73.52 \pm 1.62 km s^{-1} Mpc^{-1}$  \cite{R18} to the Planck 2018 value of $H_0=67.27 \pm 0.60 km s^{-1} Mpc^{-1}$  \cite{planck2018}.

In order to investigate possible solutions to the Hubble constant tension a number of proposals have been made \cite{VMS1}. For example, 
in \cite{Carneiro} it was shown that the 
best-fit to current experimental results includes an
additional fourth, sterile, neutrino family  with a mass of an eV order suggested by flavour oscillations.
This would imply an additional relativistic degree of freedom ($N_{eff} = 4$) in the standard model, which may alleviate the $H_0$ tension.
Recently it was argued that GW could represent a new kind of standard ``sirens"
that will  allow for $H_0$ to be constrained in a model independent way \cite{siren}.
It is unlikely that
inhomogeneitites and cosmic variance can resolve the tension \cite{Macpherson}.
However, there are suggestions that
the emergence of spatial curvature may alleviate the tension  \cite{Bolejko18,CCCS,Brand,Macpherson,Ryan2019}.
Any definitive measurement of a non-zero spatial curvature would be crucial
in cosmology. We will revisit this later.

\newpage

\section{Problems in theoretical cosmology}

\subsection{Acceleration: dark energy}

The most fundamental questions in cosmology, perhaps, concern dark matter and dark energy,
both of which are `detected' by their gravitational interactions 
but can not be directly observed \cite{Martin19}.

Indeed, the dark energy problem is believed to be one of the major
obstacles to progress in theoretical physics
\cite{Witten2001,Steinhardt}. 
Weinberg discussed the  {\em{cosmological constant  problem}} in detail 
\cite{Weinberg1989}. Conventional quantum field theory
(QFT) predicts an enormous energy density for the vacuum. 
However, the GR equivalence principle 
asserts that all forms of mass and energy gravitate in an identical manner,
which then implies that the vacuum energy is  gravitationally equivalent to
a cosmological constant and
would consequently have a
huge effect on the  spacetime curvature.
But the observed value for the effective
cosmological constant is so very tiny (in comparison to
the  predictions of QFT) that a ``bare'' cosmological
constant, whose origin is currently not known,  is necessary to 
cancel out this enormous vacuum energy to at least $10^{-120}$.
This impossibly difficult fine-tuning problem becomes even
worse if we include higher order LQG corrections 
\cite{Padilla}.

A number of authors, including  Weinberg, have offered the opinion
that of all of the possible solutions to the dark energy problem, perhaps the most reasonable is the
anthropic bound, which is itself very controversial \cite{Weinberg1987}. 
However, another possibility is that the quantum vacuum does not gravitate. This will be true if the real gravitational theory is unimodular gravity, leading to the trace-free EFE \cite{tracefree}.


Furthermore, the  expansion of the Universe has been increasing
for the last few billion years \cite{Riess,Perlmutter}. 
Within the  paradigm of standard cosmology, it is usually proposed that this acceleration is caused by a so-called {\em{dark energy}}, which effectively 
has the same properties as a very small cosmological constant (which is a
repulsive gravitational force  in GR). 
This {\em{cosmological coincidence problem}}, which necessitates a possible explanation for why 
the particular small observed valued of the
cosmological constant currently is of a similar
magnitude to that of the  matter density in the Universe,
is an  additional problem.
In particular, it is often postulated that
dark energy is not due to a pure cosmological constant but that  dynamical models such as,
for example,  quintessence and phantom
energy scalar field models, are more reasonable. 
Alternative
explanations for these gravitational effects have been proposed 
within theories with modified gravity on large scales,
which consequently do not necessitate new forms of matter.
The possibility of an effective acceleration of the Universe 
due to backreaction 
has also been discussed.

\subsection{Acceleration: inflation} 

Inflation is a central part of modern theoretical cosmology.
The assumption of zero spatial
curvature ($k = 0$) is certainly well motivated in the standard model by inflation.

Before the development of inflation, it was already known that a scale invariant
(Harrison-Zeldovitch) power spectrum is a good fit to the data. But
its origin was mysterious and there was no convincing physical mechanism to explain it.
However, inflation naturally implies this property as a result of 
cosmological perturbations of quantum mechanical
origin. Moreover, it allows a bridge to be built between 
theoretical
considerations and actual astrophysical measurements.
One fundamental assumption of inflation is that, initially, the quantum perturbations
are placed in the vacuum state \cite{Martin}.


As noted earlier, models with a positive cosmological constant
are asymptotic at late times to the inflationary de Sitter spacetime \cite{Friedrich1986,Wald83}.
Scalar field  models with an increasing rate of (volume)
expansion are also future inflationary.
For models 
with an exponential potential, global asymptotic results can be obtained \cite{Coleybook,exppot}. Inflationary
behavior is also possible in scalar field models with a power law
potential, but typically occurs during an intermediate epoch rather than asymptotically to the future.
Local results in this case are possible, but they are difficult to obtain and this problem is 
usually studied numerically.

There are a number of fundamental questions, which include the following.  What exactly is
the conjectured inflaton?
What is the precise physical details of 
{\em{cosmic inflation}}?  If inflation is self-sustaining
due to the amplification of fluctuations in the quantum regime,
is it still taking place in some 
(distant) regions of the Universe? And, if so, 
does inflation consequently give rise to an infinite number of ``bubble universes''? 
In this case, 
under what (initial) conditions can such a {\em{multiverse}} exist?
An investigation of 
``bubble universes", in
which our own  Universe is but one of many that nucleate and grow within an
ever-expanding false vacuum, 
has been undertaken (primarily computationally).  For example, the interactions
between such bubbles were investigated
in \cite{bubbles}.

Cosmological inflation is usually taken  as a reasonable explanation for the fact
that the Universe is apparently more uniform on larger scales than is anticipated within the standard cosmology (the {\em{horizon problem}}). However, 
there are other possible explanations.
But how does inflation start? And, perhaps most importantly,
what is the generality for 
the onset of
inflation for generic spatially inhomogeneous  initial data?
We note that a rigorous formulation of this question is problematic
due to the fact that there are so many different inflationary theories and
since there are no ``natural'' conditions for the initial data.
However, any such natural initial  conditions
are expected to contain some degree  of inhomogeneity
{\footnote{Note that preliminary calculations in quantum field theory suggest that vacuum 
fluctuations could induce an enormous cosmological constant
\cite{Carlip19}.}}. Unfortunately,
such initial data does not necessarily lead to
inflation. 
Although it is known 
that large field inflation can occur for simple inhomogeneous 
initial data 
(at least for energies with substantial initial gradients 
and when the inflaton field is  on the
inflation supporting portion of the potential to begin with),
it has also been shown that small
field inflation is significantly  less robust in the presence of inhomogeneities \cite{infl} (also see \cite{bubbles} and \cite{infl2}).


\subsubsection{Alternatives to inflation}

Although inflation is the most widely acceptable mechanism for the generation of almost scale invariant 
(and nearly Gaussian adiabatic density)
fluctuations to explain the origin of structure on large scales, 
possible alternatives include GR spikes  \cite{art:ColeyLim2012}, conformal cyclic cosmology 
 \cite{Penrosebooks} and QG
fluctuations \cite{Hamber}. In particular, Penrose has argued that since inflation fails to take fully into account the huge gravitational entropy that would be associated with black holes in a generic spacetime, inflation is incredibly unlikely to start, and smooth out the universe, if its initial state is generic \cite{Penrosebooks}.  In addition, in the
approach of \cite{Hamber} results 
from non-perturbative studies of QG regarding the large 
distance
behavior of gravitational and matter two-point functions are utilized;  non-trivial scaling dimensions exist due to
a nontrivial
ultraviolet renormalization group fixed point in 4D, motivating 
an explanation for the galaxy power spectrum  based on the 
non-perturbative quantum field-theoretical treatment of GR.
Perhaps the most widely accepted alternative to inflation to obtain successful structure formation
and which is consistent with current observations \cite{beyond} is the matter bounce scenario, in which new physics resolves the cosmological singularity. 

\subsubsection{Bouncing models revisited}

Bouncing models include the ekpyrotic 
and emergent
string gas  scenarios
\cite{beyond}. 
The  {\em{ekpyrotic}} scenarios \cite{Khoury} are bouncing cosmologies which avoid the
problems of the anisotropy
and overproduction of GW in the matter bounce scenario, since the
dynamics of the 
contracting phase is governed by a matter field (e.g., a scalar
field with negative exponential potential) whose energy density increases  faster than the contribution of anisotropies.
In ekpyrotic scenarios, in which the bounce  is not necessarily symmetric, 
fluctuations on all currently observable scales 
start inside the Hubble radius at earlier times, 
leading to structure that is  formed causally 
and hence a solution of the horizon problem 
in the same way
as in standard big bang cosmology
and as in the usual matter bounce. 
But, contrary to the matter bounce scenario, during contraction the growth of 
fluctuations on super-Hubble scales is too weak to produce a scale-invariant 
spectrum from 
an initial vacuum state, leading to a subsequent blue spectrum of curvature
fluctuations and GW  \cite{beyond}.
Therefore,  initial vacuum perturbations cannot describe
the observed structures in the Universe. In addition, a negligible amplitude of GW is predicted on cosmological
scales. However,  a scale invariant
spectrum of curvature 
fluctuations can be obtained by using primordial vacuum 
fluctuations in a second
scalar field in the ekpyrotic scenario \cite{Notari}.

Another alternative to cosmological inflation is the {\em{emergent
string gas}}  scenario \cite{BrandenbergerVafa},  based
on a possible extended quasi-static period in the very early Universe dynamically dominated by a
thermal gas of fundamental strings, after which 
there is a transition to the expanding radiation phase of standard cosmology.
The thermal 
fluctuations of a gas of closed strings on a compact space with toroidal topology, which do 
not originate quantum mechanically
(unlike in most models of inflation), then produce
a scale-invariant spectrum of curvature 
fluctuations and GW. The tilt of the spectrum
of curvature 
fluctuations is predicted to be red as in inflation, but that of the GW
is slightly blue, in contrast to what is obtained in inflation.

We should note that although some of the alternatives to inflation are suggested by ideas
motivated by QG, it is also of interest to know whether inflation occurs naturally
within QG. We will discuss this later.

\subsection{The physics horizon and Synge's g-method} 

\noindent
{\em{The Physics horizon:}}
The basic problem as regards inflation and any attempts to model what happened at earlier times in the history of the Universe is that we run into the \textit{physics horizon}: we simply do not know what the relevant physics was at those early times. The reason is that we cannot construct particle colliders that extend to such high energies. 
Thus we are forced either to extrapolate tested physics at lower energies to these higher energies, with the outcome depending on what aspect of lower energy physics we decide to extrapolate (because we believe it is more fundamental than other aspects), or to make a phenomenological model of the relevant physics.

\noindent
{\em{Synge's g-method:}}
A very common phenomenological method used is \textit{Synge's g-method}: running the 
EFE backwards \cite{EllMaaMac}. That is, in eqn. (\ref{eq:EFE}), instead of trying to solve it from right to left (given a matter source, find a metric $\textbf{g}$ that corresponds to that matter source), rather choose the metric and then find the matter source that fits. That is, select a metric $\textbf{g}$ with some desirable properties, calculate the corresponding Ricci tensor $R_{ab}$ and Einstein tensor $G_{ab}$ and then use (\ref{eq:EFE}) to find the matter tensor $T_{ab}$ so that (\ref{eq:EFE}) is identically satisfied, and \textit{voila!} we have an exact solution of the EFE that has the desired geometric properties. No differential equations have to be solved. The logic is: via (\ref{eq:EFE}), 
\begin{equation}\label{eq:synge}
\{g_{ab}\} \Rightarrow \{R_{ab}\} \Rightarrow \{G_{ab}\} \Rightarrow \{T_{ab}\}.
\end{equation} 
One classic example is choosing an inflationary scale factor $a(t)$ that leads to structure formation in the early Universe that agrees with observations. We can then run the EFE backward as in (\ref{eq:synge}) to find a potential $V(\phi)$ for an effective scalar field $\phi$ that will give the desired evolution $a(t)$. It is a theorem that one almost always can find such a potential \cite{Ell_Mads}, essentially because the energy momentum conservation equations are in that case equivalent to the Klein-Gordon equation for the field $\phi$; but there is no real physics behind claims of the existence of such a scalar field. It has not been related to any matter or field that has been demonstrated to exist in any other context.  

\subsection{Dynamical behaviour of cosmological solutions} 
The dynamical
laws governing the evolution of the universe are the classical EFEs.
It is of interest to study
exact cosmological solutions and especially spatially inhomogeneous cosmologies \cite{krameretal},
and their qualitative and  numerical behaviour.
Dynamical systems representations of the evolution of cosmological solutions are very useful  \cite{Collinsandellis,WE}.  
In particular, it is of interest to extend stability results to 
the study of cosmological models with matter and in
the case of a non-zero cosmological constant
\cite{Dotti}.

\subsubsection{Stability of cosmological solutions}

This concerns the question of 
whether the evolution of the
EFE under small perturbations is qualitatively similar to the evolution of the underlying 
exact cosmological solution  (e.g., by including small-scale fluctuations).
This problem involves the investigation of the (late time)  behavior of a complex set of partial differential equations about a specific cosmological solution  \cite{AnderssonMoncrief}.
The asymptotic behaviour of solutions in cosmology was reviewed in \cite{WE}.

We note that for a  vanishing cosmological constant and matter that satisfies the usual energy conditions, spatially homogeneous spacetimes of (general) Bianchi type IX recollapse and consequently do not expand for ever.
This result is formalized in the so-called
“closed universe recollapse conjecture” \cite{BarrowTipler}, which
was proven in  \cite{LinWald}.
However,  Bianchi type IX spacetimes need not recollapse in the case that a positive 
cosmological constant is present.
The study of the stability of de Sitter spacetime for generic initial data is very important,  
particularly within the context of  inflation (although, as noted earlier, precise statements concerning the generality of inflation are problematic).
 
\subsubsection{Stability of de Sitter spacetime} 

A stability  result
for de Sitter spacetime (vacuum and a positive cosmological constant) for small generic initial data
was proven in \cite{Friedrich1986}. Therefore,
de Sitter spacetime is a local attractor for expanding cosmologies
containing a positive cosmological constant. 
In addition, it was proven that any expanding spatially homogeneous model
(in which the matter obeys the strong and dominant energy conditions) that does not  recollapse is future asymptotic to an isotropic de Sitter spacetime
\cite{Wald83}. 
This so-called
``cosmic no hair'' theorem
is independent of the particular
matter fields present. 
The remaining  question is whether general, initially expanding,
cosmological solutions corresponding
to initial data for the EFE with a positive cosmological constant
and physical matter exist globally in time.
It is known that this is indeed the case for a variety of
matter models (utilizing the methods of \cite{Rendall95}).
Global stability results have also been proven
for inflationary models with a scalar
field with an exponential potential \cite{Coleybook,exppot}.
It is, of course, of considerable interest to investigate the cosmic no--hair theorem 
in the inhomogeneous case. 
A number of partial results are known in the case of a positive cosmological constant \cite{Jensen}.

The possible quantum instability of de Sitter spacetime has also been investigated. 
In a semi-classical analysis of backreaction in an expanding universe with a conformally coupled scalar 
field and a cosmological constant, it was advocated that
de Sitter spacetime is unstable
to quantum corrections and might, in fact, decay. In principle, this could consequently provide a
mechanism that might alleviate the cosmological constant problem and also, perhaps, the fine-tuning problems that occur for the very 
flat inflationary potentials that are necessitated by observations.

In particular, it has been suggested that de Sitter
spacetime is unstable due to infrared effects, in that the backreaction of super-Hubble scale GW could contribute  negatively to the
effective cosmological constant and thereby cause the latter to
diminish. Indeed,  from  an investigation of
the backreaction effect of long wavelength cosmological
perturbations it was found that
at one LQG order super-Hubble cosmological perturbations
do give rise to a negative contribution to the cosmological constant 
\cite{infra}.
It has consequently been proposed that this backreaction could then lead
to a late time scaling solution for which the contribution
of the cosmological constant tracks the contribution of the
matter to the total energy density; that is, the cosmological constant obtains a negative contribution from infrared fluctuations whose magnitude increases with time \cite{Brand}.

\subsubsection{The nature of cosmological singularities}

Although the singularity theorems imply that singularities occur generally in 
GR, they say very little about their nature  \cite{Senovilla2012}. For example,
singularities can occur in tilted Bianchi cosmologies in which all of the scalar quantities remain finite \cite{EllisandKing}. However, such cosmological models are likely not generic.
Belinskii, Khalatnikov and Lifshitz (BKL) \cite{art:LK63} 
have conjectured that within GR, and for a generic inhomogeneous cosmology, the approach to the
spacelike singularity into the past is vacuum dominated, local and oscillatory, obeying the
the so-called BKL or mixmaster dynamics. In particular, 
due to the non-linearity of the
EFE, if the matter is not an effective  massless scalar field,
then sufficiently close to the singularity {\em{all matter terms can be neglected}} in
the FE relative to the dynamical anisotropy.
BKL have confirmed
that the assumptions they utilized are consistent with the EFE. However, 
that doesn't imply that their assumptions are always valid in  general  situations of physical interest.
Numerical simulations have recently been used to verify the BKL dynamics in
special classes of spacetimes \cite{Berger,DavidG}. Rigorous mathematical results on the 
dynamical behaviour of Bianchi type VIII and IX cosmological models have also been presented \cite{bianchi}.

Up to now there have essentially been three main approaches to investigate the structure of generic singularities,
including the original heuristic BKL metric approach and the so-called Hamiltonian approach.
The dynamical
systems approach \cite{WE}, in which the
EFE are reformulated as a scale invariant asymptotically regularized dynamical system (i.e., a first order system of autonomous ordinary or partial differential equations) in the approach towards a generic spacelike singularity, allows for a
more mathematically
rigorous study of cosmological singularities.
A dynamical systems formulation of the EFE (in which
no symmetries were assumed a priori) was presented in \cite{Uggla03}, which resulted in
a detailed description of the generic attractor, precisely formulated conjectures
concerning the asymptotic dynamical behavior toward a generic spacelike singularity, and a 
well-defined framework for the numerical study of cosmological singularities \cite{Andersson}.
It should be noted that these studies assume that the singularity is spacelike, but there is no reason that this has to be so (this is not, in fact, generic).
The effect of GR spikes on the BKL dynamics and on the initial cosmological singularity was reviewed in \cite{MathGR}.

\subsubsection{Isotropic singularity} 

Penrose  \cite{Penrosebooks} has utilized entropy considerations to motivate the ``Weyl curvature hypothesis'' that asserts
that on approach to an initial cosmological singularity the Weyl curvature tensor should tend to
zero or at least remain bounded 
(this conjecture subsequently led to the conformal cyclic cosmology proposal).
It is difficult to represent this proposal 
mathematically but the clearly formulated
geometric condition presented in \cite{Goode},
that the conformal structure should remain regular at the singularity, is
closely related to the original Penrose proposal.
Such singularities are called isotropic or conformal singularities.
It is  known \cite{Claudel} that
solutions of the EFE for a radiation perfect fluid that admit an isotropic singularity are uniquely characterized by particular free
data specified at the singularity. The required data is essentially
half as much as the data necessary in the case of a regular Cauchy hypersurface.
This result was generalized to the case of a perfect fluid with a linear equation of state\cite{Anguige}, 
and can be further extended to more general matter models (e.g., more general fluids 
and a collisionless gas of 
massless particles) \cite{Rendall2002}.

As noted earlier, we do not aim to discuss
alternative theories of gravity in this review. 
However, it is of cosmological interest to determine whether
isotropic singularities are typical in any modified theories of gravity.
For example, the past stability of the isotropic FLRW vacuum solution,  on 
approach to an initial cosmological singularity, in the class of theories of gravity containing 
higher--order 
curvature terms in the GR Lagrangian, has been investigated \cite{Middleton}.
In particular, a special
isotropic vacuum solution was found to exist, which behaves like a radiative FLRW model, that is
past stable to small anisotropies and inhomogeneities (which is not the case in GR).
Exact solutions with an isotropic singularity for specific values of the perfect fluid equation of state parameter have also been  obtained 
in a higher dimensional flat anisotropic Universe 
in Gauss-Bonnet gravity 
\cite{Kirnos}. 
A number of simplistic cosmological solutions of theories of gravity
containing a quadratic Ricci curvature term in the Einstein-Hilbert Lagrangian  
have also been investigated
\cite{BarrowHervik}.

\newpage

\section{Problems in physical cosmology}

The  predicted distribution of {\em{dark matter}} in the Universe is based on observations of
galaxy rotation curves,
nucleosynthesis estimates and computations of structure formation \cite{Freese}. 
The nature of the missing  dark matter
is not yet known (e.g., whether it is due to a particle or whether the dark
matter phenomena is not characterized by any type of matter but rather by a modification of GR).
But it is, in general, anticipated that this particular problem will be explained within conventional physics. 
More recently primordial black holes have been invoked to explain the missing
dark matter and  to alleviate some
of the problems associated with the CDM scenario (see later) \cite{BernardCarr}.

\subsection{Origin of structure}

The CMB anisotropies and structure observed on large angular scales
are  computed using linear 
perturbations about the standard
background cosmological model. However, 
such large scale structure
could never have been in causal contact 
within conventional cosmology and hence its origin cannot be
explained by it without invoking inflation.
In general, the testable predictions of inflationary models are
scale-invariant and nearly Gaussian adiabatic density 
fluctuations and almost, but not exactly, a scale-invariant stochastic background of relic 
GW. However, and as noted earlier,
possible alternatives to inflation to obtain successful structure formation
consistent with current observations \cite{beyond} exist, including
the popular matter bounce cosmologies. 

\subsubsection{Large scale structure of the Universe}

In the standard cosmology it is assumed that cosmic structure at sufficiently large scales
grew out of small initial fluctuations at early times, and  we
can study their evolution within (cosmological)  linear perturbation theory (LPT) \cite{Durrer1996}.
We assume that on large scales there is a well defined mean density and on
intermediate scales, the density differs little from it.
This is a highly non-trivial assumption, which is perhaps justified by the isotropy of the
CMB. It is usual to use a fluid model for matter and a kinetic theory model for radiation.

At late times and sufficiently small scales fluctuations of the cosmic density are not
small. The density inside a galaxy is about two orders of magnitude greater than the mean density
of the Universe, and
LPT is then not adequate to study
structure formation on galaxy-cluster scales of a few Mpc and less.
It is necessary to
treat clustering non-linearly using N-body simulations.
Since this is mainly relevant on scales much smaller than the Hubble scale, 
it has usually been studied in the past with non-relativistic N-body simulations. 
On intermediate to small scales, density perturbations can become large.
Inside a galaxy they are small, and even inside a galaxy cluster the motion of
galaxies is essentially decoupled from the Hubble flow (i.e., clusters do not expand).  Therefore, the gravitational potential of a galaxy remains small, and in
the Newtonian (longitudinal) gauge, metric perturbations remain small.
In the past, this together with the smallness of peculiar velocities has been used to
argue that Newtonian N-body simulations are sufficient. 

In the adiabatic case, the last scattering surface is a surface of constant baryon density, so the observed CMB fluctuations do not represent density fluctuations, as is often stated
\cite{Durrer}. Thus, in standard perturbation theory language, this shows that in the uniform density gauge (which for adiabatic perturbation is the same as the uniform temperature gauge) the density fluctuations are given exactly by the redshift fluctuations. In the non-adiabatic case this will no longer be true. 
The main shortcoming of the conventional analysis is, of course, the instantaneous recombination approximation (accurate to a few percent only for multipoles with $\ell<100$); to go beyond this one has to use a Boltzmann approach~\cite{Durrer} (although nothing changes conceptually).
Also, in principle we cannot
neglect radiation or 
neutrino (even massive) velocities. In addition, 
Newtonian simulations only consider 1 (of in general 6) degrees of freedom, and
observations are made on the relativistic, perturbed light cone. Hence relativistic calculations are needed.

\subsubsection{Perturbation theory}

The complexity of the distribution
of the actual matter and energy in our observed Universe, consisting of stars and galaxies
that form clusters and superclusters of galaxies across a broad range of scales,
cannot be described within the standard spatially homogeneous  model. To do
this we must to be able to describe spatial inhomogeneity and anisotropy using a
perturbative approach starting from the uniform
FLRW model as a background solution \cite{MalikWands}.
The perturbations
“live” on the  four-dimensional background spacetime, which is split into 
three-dimensional spatial hypersurfaces
utilizing a (1+3) decomposition. Within the standard cosmology a flat background spatial metric ($k=0$)
in LPT is assumed, which is consistent
with current observations. For generalisations to spatially hyperbolic
or spherical FLRW models see, e.g., \cite{KodamaSasaki}.

The introduction of a spatially homogeneous background spacetime to describe the inhomogeneous
Universe leads to an ambiguity in the choice of coordinates. 
Selecting a set of coordinates in
the (real) inhomogeneous Universe, which will then be described by an FLRW model
plus perturbations, essentially amounts to the assignation of a mapping between spacetime points
in the inhomogeneous Universe and the spatially homogeneous background model. The
freedom in this selection is the gauge freedom, or gauge problem, in GR
perturbation theory. 
Either the gauge freedom must be handled very carefully  by delineating what freedom remains at each stage of coordinate specialisation \cite{SachsWolfe}, by using gauge covariant variables \cite{Bardeen}, or utilizing  1+3 gauge invariant and covariant variables \cite{gaugeinv}.

Indeed, gauge-invariant variables are widely utilized since they constitute a
theoretically effective
way to extract predictions from a gravitational field theory applied to the Universe for
large-scale linear evolution \cite{KodamaSasaki}. In addition,
by using gauge-invariant variables the analysis is reduced to the study of only three decoupled second order
ordinary differential equations, and they represent physical quantities
that can be immediately connected to observations. In the review \cite{MalikWands} the focus was on how to construct a variety of gauge invariant
variables to deal with perturbations in different cosmological models
at first order and beyond.
Most work to date has been done only to linear order where the perturbations
obey linear FE.

As a theoretical application the origin of primordial curvature and isocurvature
perturbations from field perturbations during inflation in the very early Universe can be considered.
LPT allows the primordial spectra to be related to quantum
fluctuations in the metric and matter fields at considerably higher energies. In the
most simple single field inflationary models it is, in fact, possible to equate the primordial
density perturbation with the curvature perturbation
during inflation, which essentially remains constant on very large scales for adiabatic density
perturbations. 
The observed power spectrum of primordial perturbations revealed by the CMB
and LSS is thus a powerful probe of
inflationary models of the very early Universe.

The outstanding problems within LPT are mostly technical issues and, in particular,  include the 
important questions of the  physical cut off to the short and long wavelength modes  
and the convergence of the perturbations (and hence the validity of the perturbative approach itself).

\subsubsection{Non-linear perturbations} 

The new frontier in cosmological perturbation theory is the investigation of
non-linear primordial perturbations, at second-order and beyond.
Although the simple evolution equations  obtained at linear order can be  extended 
to non-linear order \cite{MalikWands}, the non-linearity
of the EFE becomes evident and consequently
the resulting  definitions of gauge invariant quantities at second order clearly become more
complicated than those at first order.
Recently, perturbations  at
second order  \cite{MalikWandssecond} and, more generally, non perturbative effects have been studied,
where there are certainly more foundational problems.

Perturbative methods allow quantitative statements but have
limited domains of validity.
Recently, several groups have started to develop
relativistic simulations \cite{Computational}.
Numerical relativistic N-body simulations are a unique tool to study more realistic scenarios, and
appear to compare well to numerical relativity fluid
simulations  \cite{EastWojtak}. However, assumptions are still made that need to verified.
In particular, care must be taken in applying Newtonian intuition to GR.
For example, \cite{Adamek} do not solve the full EFE
and use the fact that the gravitational potential is very small, but spatial derivatives, 
and second derivatives, are not small.
Therefore, when computing the Einstein tensor they go only to first order in the gravitational
potentials and their time derivatives (but also include quadratic terms of first spatial
derivatives and all orders for second spatial derivatives). 

New qualitatively effects occur beyond linear order. The non-linearity of the
FE inevitably leads to mixing between scalar, vector and tensor
modes and the existence of primordial density perturbations consequently
generate vector and tensor modes.
Non-linearities then permit additional information to be determined from the primordial
perturbations. A lot of effort is currently being devoted to the investigation of higher order
correlations (and issues of  gauge dependence).
Non-Gaussianity in the primordial density
perturbation distribution would uncover interactions beyond the linear theory. Such interactions
are minimal (suppressed by slow-roll parameters) in the simplest
single field inflation models, so any detection of primordial non-Gaussianity
would cause a major reassessment
about our knowledge of the very early Universe.
In principle, this approach can be easily extended to higher-orders, although large primordial non-Gaussianity 
is expected to dominate over non-linearity in the transfer functions.

However, cosmological perturbation theory  based on a cosmological $1+3$ split
is ill-suited to address
important questions concerning  non-linear dynamics or to evaluate the viability of scenarios
based on classical modifications of GR. 
A new formulation  of a fully non-perturbative approach has been advocated \cite{Ijjas},
along with a gauge fixing protocol that enables the study of these issues (and especially the linear mode stability in spatially homogeneous and nearly homogeneous backgrounds)
in a wide range
of cosmological scenarios, based on a method that has been successful
in analyzing dynamical systems in mathematical and numerical GR based on the generalized harmonic formulation
of the EFE.

\subsubsection{Non-linear regime}

At the non-linear order a variety of different effects come into play, including
gravitational lensing of the source by the intervening
matter and the fact that redshift is affected by
peculiar motion, both of which have relatively simple
Newtonian counterparts. But there are a host of complicated
relativistic corrections once light propagation is
worked out in more detail. There are selection effects
too: we are much more likely to observe sources in halos,
some objects are obscured from view by bright clusters,
and so on.

Within the context of perturbation theory
it is relatively easy to predict the expectation value of
the bias in the Hubble diagram for a random direction \cite{Fleury17}.
The full second-order correction to the distance-redshift relation has
been calculated  within cosmological
perturbation theory, yielding the observed redshift and the lensing magnification to 
second order appropriate  for most investigations of dark energy models
\cite{Umeh14}.
These results were used in \cite{ClarksonUmehMaartens}  to calculate the impact of second-order perturbations
on the measurement of the distance to the last-scattering surface, where
relativistic effects can lead to significantly biased measurements of the cosmological parameters
at the sub-percent to percent level if they are neglected.

The somewhat unexpected percent level
amplitude of this correction was discussed in \cite{Bonvin15}, but the focus therein was on
on the effect of gravitational lensing only
and thus did not consider the perturbations of the observed redshift, notably due to peculiar
velocities,
which  can lead to a further bias in parameter 
estimation. In addition, 
\cite{Ben-Dayan13}  noted that the notion of average is
adapted to the observation of the Hubble diagram and  may differ from the most common
angular or ensemble averages, and   suggested  a possible non-perturbative way for computing the effects of inhomogeneities on observations based on light-like 
signals using the geodesic light-cone  gauge to 
explicitly solve the geodetic-deviation equation.

In order to
comprehensively address the issue of the bias of the distance-redshift relation,
previous work was improved upon by fully
evaluating the effect of second-order perturbations on the Hubble diagram  \cite{Fleury17}.
In particular, 
the notion of average which affects bias in observations of the 
Hubble diagram for inhomogeneity of the 
Universe was carefully derived, and  its bias at second-order in cosmological perturbations
was  calculated. 
It was found that this bias considerably affects direct estimations of the evolution of the
cosmological parameters, and particularly the equation of state
of dark-energy. 
Despite the fact that the bias effects can reach the percent level 
on some parameters, errors in the standard inference of cosmological 
parameters remain less than the uncertainties in observations \cite{Fleury17}.

In further work \cite{Adamek18}, a non-perturbative and
fully relativistic numerical calculation of the observed luminosity
distance and redshift for a realistic cosmological
source catalog in a standard cosmology was undertaken to investigate the bias and scatter, mainly due to gravitational lensing and peculiar velocities,
in the presence of cosmic structures.
The numerical experiments provide conclusive evidence
that the non-linear  relativistic evolution of inhomogeneities,
once consistently combined with the kinematics of light
propagation on the inhomogeneous spacetime geometry,
does not lead to an unexpectedly large bias on the
distance-redshift correlation in an ensemble of cosmological
sources. However, inhomogeneities introduce
a significant non-Gaussian scatter that can give a large
standard error on the mean when only a small sample of
sources is available. But even for large, high-quality  supernovae samples
this scatter can bias the inferred cosmological parameters
at the percent level \cite{Adamek18}.

It was argued in \cite{BenDayan14}, using a fully relativistic treatment, that cosmic variance 
(i.e.,  the effects of the local structures such as galaxy clusters  
and voids) is of a similar order of magnitude to current 
observational errors and consequently needs to be taken into 
consideration in local measurements of the Hubble expansion 
rate within the standard cosmology.
In addition,  the constraint equation relating metric and density perturbations in GR is 
inherently non-linear, and leads to an effective  and intrinsic non-Gaussianity in the  large-scale dark matter density field on 
large scales (even when the primordial metric perturbation is itself Gaussian) \cite{Bartolo}. 

\subsubsection{Non-Gaussianities} 

In standard cosmology, the primordial perturbations corresponding to
the seeds for the LSS are selected from a Gaussian distribution with random phases,
justified primarily from the fact
that primordial non-Gaussianity (PNG) has not yet been observed and also theoretically (e.g., the central limit theorem); thus a Gaussian random field constitutes a satisfactory representation
of the properties of density fluctuations.
However, any deviation from perfect Gaussianity will, in principle, reveal important information on
the early Universe, and an investigation of PNG is especially relevant if these initial conditions were generated by some dynamical process such as, for example, inflation. 
In particular, a direct measurement of non-Gaussianity would permit us to move beyond the free-field
limit, yielding important information about the degrees of freedom, the possible symmetries
and the interactions characterizing the inflationary action. 
The current status of the modelling of, and the searching for,  PNG  of cosmological perturbations was reviewed in  \cite{Celoria}.

In order to evaluate PNG from the early Universe to the present time, it is necessary
to self-consistently calculate non-Gaussianity during inflation.
We must then evolve scalar and tensor perturbations to second order outside the horizon, matching conserved second-order gauge-invariant variables to their values at
the end of inflation (appropriately taking into account reheating).
Finally, we need to investigate the evolution of the perturbations after they re-entered the
Hubble radius, by computing the second-order radiation transfer function 
and matter transfer function
for the CMB and LSS, respectively.
Although these calculations are very complicated, PNG represents an important tool to probe
fundamental physics  during inflation at energies from the grand unified scale, since different
inflationary models predict different amplitudes and shapes of the bispectrum, 
which complements the search for primordial gravitational-waves
(PGW) (via a stochastic GW background).

The Planck satellite has produced good measurements of
higher-order CMB correlations, resulting in considerable stringent constraints on PNG.
The latest data regarding non-Gaussianity tested the local, equilateral, orthogonal (and various other) shapes for the bispectrum and led to new constraints on the
primordial trispectrum parameter \cite{planck2018}.
The most extreme possibilities have been excluded by
CMB and LSS observations,  and now primarily
the detection of (or
constraints from) mild or weak deviations from primordial Gaussian initial conditions are sought,
characterized by a small parameter, $f_{NL}$, compatible  with observations. 
Even though the sensitivity is not comparable to CMB data \cite{planck2018},
the bispectra for 
redshift catalogues can be determined (e.g., the three-point correlation functions for the WiggleZ and BAO
spectroscopic surveys) \cite{GilMarin}, 
and interesting observational bounds on the
local $f_{NL}$ from current constraints on the power spectrum can be obtained (see  \cite{Celoria} and references within). Neglecting 
complications arising from the breaking of statistical isotropy (such as sky-cut, noise, etc.) the
procedure is, in general, to fit the theoretical bispectrum template, and
$f_{NL}$ is found to be approximately 0.01 in generic inflation \cite{Gauss}.

PNG is certainly the best way of practically investigating
the only guaranteed prediction of
inflation \cite{SilkLimits}. 
Indeed, even though standard models of slow-roll inflation only predict tiny deviations from Gaussianity (consistent with the Planck results), specific oscillatory PNG features can
be indicative of particular string-theory models.
Therefore, the search for PNG is of interest for theoretically well-motivated
models of inflation and the Planck results  can potentially severely
constrain a variety of
classes of inflationary models beyond the simplest paradigm.
However, only the failure to find any such evidence for PNG
can falsify inflation.

There are some outstanding issues regarding non-Gaussianity \cite{Celoria}. First,
it has been  argued that the  consistency relation  is certainly not observable for
single field inflation since, in the strictly squeezed limit, this term can be gauged away by an
appropriate coordinate tranformation (so that the only residual term is
proportional to the same order of the amplitude of tensor modes).
Second, in the non-linear evolution of the matter
perturbations in GR the second order dark matter dynamics leads to
post-Newtonian-like contributions which mimic local PNG. 
A recent estimate of the effective non-Gaussianity due to GR light cone
effects comparable to a PNG signal were discussed in \cite{Celoria}, which would
correspond in the comoving gauge to an $f_{NL}$ in the pure squeezed limit.
Therefore, such a GR  PNG signature may not be detectable via any cosmological observables.

\subsubsection{Simulations and post-Newtonian cosmological perturbations}

There has been a lot of  recent interest
in testing the validity of GR using cosmological observables related to structure formation.  
Since the physics
involved in horizon-sized cosmological perturbations is
quite different to that which occurs on smaller scales,
where galaxies and clusters of galaxies are present, this is challenging.
LPT  \cite{MalikWands} is not
suitable for investigating gravitational fields associated with
structures that have highly non-linear
density contrasts (which necessarily have to be small in order for the perturbative expansion to be well defined). 
GR numerical simulations using, for example, the \textit{gevolution} code developed by Adamek, Durrer and co-workers \cite{Adamek}, have proven to be an important new tool for studying
structure formation. Targeted fully relativistic non-linear
simulations with an evolving non-zero spatial curvature have also been developed \cite{Bolejko18}.

Alternatively, 2-parameter post-Newtonian cosmological perturbation schemes have been proposed \cite{Goldberg}. 
Indeed, recent progress \cite{Sanghai} has been made in applying the techniques from post-Newtonian expansions
of the gravitational FE
into cosmology in the presence of
highly non-linear structures 
to relate the functions
that parameterize gravity on non-linear scales to those that parameterize it on very large
scales. This so-called parameterized
post-Newtonian cosmology (PPNC)
has been used to analyse alternative theories of gravity \cite{Sanghai}.
This was achieved  by simultaneously expanding all of the relevant equations in terms of two parameters; the first associated with the expansion parameter of LPT, and the second characterizing 
the order-of-smallness from post-Newtonian theory
\cite{Goldberg}. An alternative 
Lagrangian-coordinates based  approximation scheme to provide a unified treatment for the two 
leading-order regimes was presented in \cite{Rampf}.

\subsection{Black holes and gravitational waves}

\subsubsection{Gravitational waves} 

Recent progress in numerical GR has allowed for a detailed investigation of the  
collision of two compact objects (such as, e.g., black 
holes and neutron stars). In such a 
violent inspiral
an enormous amount of gravitational radiation is emitted. The  
detection and subsequent analysis of the  gravitational wave (GW) signals produced 
by black hole mergers necessitate extremely accurate
theoretical predictions that can be utilized as 
template waveforms that can then be  used to cross-correlate with the output of 
GW detectors. This is, of course, of fundamental import in view
of the recent LIGO observations \cite{LIGO}. 
Indeed, such an analysis led to the 
direct detection of GW by the LIGO-Virgo collaboration \cite{LIGO2}. 
To a large extent the numerical problem has been solved in 
the case of a black-hole merger, although
the relatively simple properties of  the  two-body non-linear gravity waveforms \cite{YangPaschalidis}
have not been fully understood mathematically. There is also the recent binary neutron star merger event,
which is much more difficult to model within GR.
There are a number of open problems, particularly concerning the physical nature of the recently observed 
merger events \cite{Barack}. 
GW astronomy will potentially play an increasingly important role within cosmology \cite{generalGWREFS}. For example, there is a promise that they will allow very good direct estimates of the distance of colliding black holes, avoiding the need for the usual cosmic distance ladder.

\subsubsection{Primordial gravitational waves} 

Primordial GW (PGW)  add to the relativistic degrees of freedom of the cosmological fluid. Any change in the particle physics content, perhaps due to a change of phase or freeze-out of a species, will leave a characteristic imprint on an otherwise featureless spectrum of PGW.
The existence of a stochastic PGW
background at a detectable level would then probe new physics beyond the standard 
cosmological model,
and this may be possible with the
Laser Interferometer Space Antenna (LISA) 
\cite{LISA}.

Recently, a class
of early-Universe scenarios 
has been theoretically identified 
which produce a strongly amplified,
blue-tilted spectrum  of GW \cite{Caldwell}.
Detection of GW over a broad range
of frequencies can provide important information concerning the
underlying source \cite{Caldwell}, and
also may well be of relevance for the spectrum of GW
emitted by  other
scaling sources. In addition,
a population of massive  primordial black holes (PBHs) would be anticipated to generate a stochastic background of GW \cite{Carr80}, regardless of whether they form binaries or not.
The focus is usually on the GW generated by either stellar black holes (observable by LIGO) or supermassive black holes (observable by LISA). However,
with an extended PBH mass function, the GW background ought to
encompass both of these limits and also every intermediate frequency.
Many  supermassive black holes are in binary pairs that orbit together
and eventually merge, emitting GW in the process. The LISA detection window includes
mergers of black holes in the mass range of $10^4 −- 10^7$ solar masses \cite{Amaro}.
Due to the possibility that the coalescing black
holes observed by LIGO  \cite{LIGO2} could be of primordial origin, 
black holes in the intermediate mass range of $10 - 10^3$ solar masses
are of particular interest since such  PBHs might contribute to 
the dark matter (see below).

The primary goal of CMB observations is the polarization signal
induced by GW at the start of inflation. There is a considerable effort underway
to obtain stricter limits on the tensor-to-scalar ratio, r, the quantitative measure of the ratio
of the primordial amplitude of the B-mode (or shearing) polarization component due
to GW to the scalar (or compressive) mode of CMB temperature fluctuations
associated with the density fluctuations that seeded structure formation. 
While PGW have not yet been detected, the upper limit on r
from the BICEP2/Keck CMB polarization experiments  \cite{planck2018} (in conjunction with Planck temperature measurements and other data) is less than or equal to approximately 
$0.07$ at the 95\% confidence level. However,
the tensor amplitude predicted
depends on the (fourth power of the) energy scale of inflation, and so the primordial polarization signal could, in principle, be unmeasurably small \cite{SilkLimits}.

\subsubsection{Primordial black holes}

The possibility of $10 - 10^3$ solar mass objects is of
particular interest in view of the recent detection of black-hole mergers
by LIGO which has, in particular, revitalized the interest in stellar mass black holes of around
thirty solar masses (which are larger than initially expected) \cite{LIGO2}, and especially 
non-evaporating  primordial black holes (PBHs).
In particular, it has been suggested that massive PBHs could provide the dark
matter \cite{Carr16} or the supermassive black holes
which reside in galactic nuclei and power quasars \cite{CarrSilk}.


The most natural mechanism for PBH formation involves the collapse of primordial inhomogeneities, such as might arise from inflation (or spontaneously at some kind of phase transition).
Interest in PBH increased due to the discovery that black holes radiate \cite{HawkingNature}, since only PBH could
be small enough for this to be relevant cosmologically. Indeed, 
evaporating PBHs have been invoked to explain several cosmological features \cite{BernardCarr}.
Since it was initially believed that PBHs would grow as fast
as the Universe during the radiation-dominated era and consequently attain a huge mass  by
the present time, it was thought that PBH never
formed and could thus be excluded. However, such an argument is essentially Newtonian and neglects the cosmological expansion, and in   \cite{CarrHawking} it was shown that there is no self-similar solution in which a black hole can grow as fast as the Universe. 
Therefore,  once formed, their contribution to the dark matter
of the Universe grows with time (the mass of non-evaporating PBH is unchanged
after formation and can only grow if they accrete matter) \cite{DolgovSilk}.

PBH would have the particle
horizon mass at formation and could form 
as early as the Planck epoch, when QG forces are comparable to gravitational forces that at later epochs are far too weak on particle scales. However, as
the Universe expands and cools, tiny black holes of Planck mass all quickly disappear.
More massive black holes live longer and should survive until today as early Universe
relics \cite{CarrHawking}. 
Attention has consequently shifted to larger PBHs, which are
unaffected by Hawking radiation. Such PBHs might have important cosmological consequences.


Perhaps the most exciting possibility is that 
PBH larger than $10^3$ solar masses 
could provide the dark matter which comprises 
25\% of the critical density \cite{BernardCarr}. 
\footnote{An alternative to PBHs includes
persistent (or ``pre-big-bang") black holes occurring in bouncing cosmologies 
\cite{CarrColey}.}
Since PBHs formed in the
radiation-dominated era, they are not subject to the well-known cosmological
nucleosynthesis constraint that baryons can contribute at most 5\% to the critical density. PBH should thus be classified as non-baryonic and behave like any
other form of cold dark matter (CDM).
The subject has consequently become very popular
and non-evaporating PBHs may turn out to play a more important cosmological role than evaporating
ones. 

PBHs could provide the dark matter
but a number of constraints restrict their possible mass ranges  \cite{Carr16},
including those arising from gravitational microlensing, but PBHs at a level of $10\%$ of the dark matter are
still possible over a wide range of masses. 
The PBH density might
be much less than the dark matter density,
but the PBHs are not
necessarily required to provide all of the dark matter \cite{BernardCarr}. 
For intermediate mass black holes of 
$10^3$ solar masses a dark matter mass fraction of only $0.1\%$ still allows for important consequences
for structure formation. 
Cosmological structures could be generated either
individually through a `seed' effect or collectively through the `Poisson'
effect (fluctuations in the  black hole population generates an initial
density perturbation for PBH dark matter), consequently alleviating some of the 
possible problems associated with the standard
CDM scenario (even when they may only contribute a small portion of the dark matter
density). Both mechanisms for generating
fluctuations then amplify through gravitational instability to bind massive regions \cite{CarrSilk} and have been considered as
either alternatives or in conjunction with other CDM scenarios.

\subsection{Effects of structure on observations: Gravitational lensing}

A particularly important  cosmological question is whether gravitational
lensing significantly alters the distance-redshift relation $D(z)$ 
to the CMB last scattering surface
or the mean flux density of sources.
Any such $D(z)$  bias could change CMB cosmology, and the corresponding bias
in the  mean flux density could alter supernova cosmology.

In spatially homogeneous and isotropic cosmologies the ratio between the proper size of
a source and the  angular diameter distance 
is a function of redshift only.
In an inhomogeneous Universe, 
lensing by intervening metric fluctuations
can cause magnification of the 
angular size, with a corresponding change of flux density, since surface brightness is not affected by gravitational lensing.
Therefore, the apparent distance to objects at a given redshift can effectively become a randomly
fluctuating quantity.

Using conservation of photons (i.e., flux conservation), Weinberg \cite{Wei76} argued that 
in the case of transparent lenses there is no mean flux density amplification, so
that the uniform universe formula for $D(z)$ remains unchanged
(where the averaging is over sources, and the result relies 
on the implicit assumption that the area of a constant-$z$
surface is unaffected by gravitational lensing).
This issue has recently been revisited  \cite{KP},
and it was argued that in an ensemble averaged (and more appropriate cosmological) sense, the perturbation to
the area of a surface of constant redshift is in reality a very small
(approximately  one part in one million) effect, supporting
Weinberg's argument and
validating the usual treatment of gravitational lensing in the analysis of CMB anisotropies.

However, Weinberg's argument
regarding the mean flux density appears to contradict 
well-known theorems of gravitational lensing, such as the
{\em focusing theorem}. 
Non-linear relativistic perturbation theory to second order
indicates that there is  bias in 
the area of a surface of constant redshift and in the mean distance to the 
CMB last scattering surface. 
Indeed, a lot of investigations of  gravitational lensing
continue to advocate significant effects in the mean. 
Bolejko \cite{Bolejko2011a} (also see references in \cite{KP}) has provided a
comprehensive review of such studies, some of which
claim large effects, some of which
obtain effects at the level of a few percent (which would still be important), while others argue that the effects are exceedingly small.
A non-vanishing
perturbation to the mean flux densities of distant sources caused by intervening structures,
at least for sources that are viewed along lines of sight that avoid mass concentrations,  effectively contradict Weinberg's result. 
Recent non-linear analysis does suggest that  non-linear effects have not been proven to be
negligible \cite{Adamek18,Fleury17,Durrer}.

\subsection{Backreaction and averaging}

Averaging in GR is a fundamental  problem within mathematical cosmology \cite{fit}. 
The cosmological FE on the largest scales are derived by averaging
or coarse graining the EFE of GR. A solution of this
problem is critical for the correct
interpretation of cosmological data \cite{BC} 
(on the largest scales the dynamical
behavior can be significantly different from the dynamics in the standard cosmology; e.g.,
the expansion rate can be greatly affected  \cite{bu00}).

First, it is of great importance to provide a rigorous mathematical definition for averaging 
(tensors on a differential manifold) in GR.
A spacetime or space volume averaging approach  must be well defined and generally covariant \cite{Av,Averaging}, and produce the structure equations for the averaged
macroscopic geometry (and 
give a prescription for the correlation functions in the macroscopic FE which
emerge in the averaging of the non-linear FE), which
do not necessarily take on precisely the same mathematical
form as the original FE \cite{Averaging}.
It is straightforward to average scalar quantities and since,
in general, a spacetime is determined entirely
by its scalar curvature invariants, a specific spacetime averaging
scheme based on scalar invariants only has been proposed \cite{Coley10}.
In addition, only scalar quantities
are (space volume) averaged within the $(1+3)$ cosmological spacetime splitting approach of
Buchert \cite{bu00}.

Although the standard FLRW
$\Lambda$CDM cosmology has, to date, been very successful in explaining all of the observational data
(up to a number of potential
anomalies and tensions \cite{tension}) it does require, as yet undetected, sources of dark energy density that currently dominate the dynamics of the Universe. 
More importantly, the actual Universe is neither isotropic
nor spatially homogeneous on small scales. 
Indeed, observations of the current late epoch Universe uncovers a
very complicated picture in which the largest gravitationally bound structures, consisting of
clusters of galaxies of different sizes, form, in turn,
``knots, filaments and sheets that thread and surround very underdense
voids'' \cite{web}. An enormous fraction of the volume of the
current Universe is, in fact, contained within voids of a single characteristic size of about $30$ megaparsecs
\cite{HV1} with an almost ``empty''  density contrast \cite{Pan11}.


In principle, a number of coarse grainings over different scales is required
to reasonably model the observed  complicated gravitationally bound large scale structures  \cite{dust}.
In standard cosmology it is implicitly taken that 
the matter distribution on the largest scale
can be modeled by an ``effective averaged out'' stress-energy tensor, regardless
of the physical details of the actual coarse graining at each scale.
However, based on the two-point
galaxy correlation function, the very smallest scale on which there can be a reasonable definition of statistical
homogeneity is $70$--$120$ megaparsecs \cite{sdb12},  and even then variations 
for the number density of galaxies 
on the order of several percent still arise in the largest possible
survey volumes \cite{h05,sl09}. It is fair to say that it is not at all clear what  the
largest scale is that matter and geometry  on smaller scales can be coarse-grained such that 
the average evolution is still an exact solution of the EFE.

A smooth macroscopic geometry (with macroscopic matter fields), applicable on cosmological scales,
is obtained after an appropriate averaging.
The coarse graining of the EFE for local inhomogeneities on
small scales can generally lead to important {\em backreaction} effects (consisting of not  just the mean cosmic variance) 
\cite{NotGW} on the average dynamics of the
Universe \cite{bu00}. 
In addition, all cosmological observations are  deduced from null geodesics (the paths of photons)
which travel enormous distances, preferentially  traversing the underdense
voids of the actual Universe. But inhomogeneities perturb curved null geodesics, so that
observed luminosity distances can be significantly affected.

A consistent approach to cosmology is consequently to 
treat GR as a {\em mesoscopic} theory, which is 
applicable only on the mesoscopic scales for which it has actually
been verified, containing a mesoscopic metric field and a mesoscopic geometry.
The effective macroscopic dynamical equations on cosmological scales are then obtained by averaging.
It had originally been hoped that such a backreaction approach might help
resolve the dark energy and dark matter problems. 
However, it now seems unlikely that backreaction can replace dark energy (although
large effects are theoretically possible from inhomogeneities and  averaging \cite{BC}).
But it can certainly affect precision cosmology at the level of 1 \% \cite{Macpherson}
and may offer a better understanding of some issues in cosmology (such as the  emergence of a homogeneity scale and
non-zero  spatial curvature due to non-linear evolution of cosmic structure). 

\subsubsection{Backreaction magnitude}

This last point is very important, and since it has 
been a source of some controversy let us summarize briefly here.
The Universe is very inhomogeneous on small scales at the present time but smooth on large scales. 
It must be remembered that density pertubations
$(\delta \rho/\rho) \simeq 10^{28}$ on Earth,  but the metric is very close to Minkowski. 
To establish the backreaction effects we need approximation methods to deal with metric perturbations   $(\delta h/h) \simeq 10^{-5}$ but second derivatives $\simeq 10^{28}$. Various approaches have been tried:
	
\begin{itemize}
\item Zalaletdinov \cite{Averaging} developed a very complex bimetric averaging formalism that can, in principle, be applied in general; the effect of such averaging on cosmological observations was estimated to be of the order of  about 1 \% \cite{CPZ}. A global
Ricci deformation flow for the metric, which is generically applicable in cosmology, was
introduced by	Carfora.
\item Buchert  \cite{bu00} developed an explicit ($1+3$) spatial averaging scheme,  although the scheme is not fully deterministic and depends on some \textit{ad hoc} phenomenological assumptions. Models based on this scheme, and particularly the ``timescape cosmology'' of  Wiltshire which utilizes time-dilation effects in voids, can predict  very large effects, and there have been
claims that the results are sufficient to explain dark energy  
\cite{BC}.  
\item There have been various approximation schemes  that have claimed that 
the backreaction effects are negligible ($\simeq 10^{-5}$), including a scheme
by Green  and  Wald  \cite{GreenWald,NotGW} which
uses distributional methods that does not involve explicit averaging.
\item  Durrer and collaborators have developed detailed second order calculations that predict percent level changes (i.e., that are sufficient to be of significance in GR precision cosmology studies), and
they (using the ``gevolution'' numerical code) and others  have subsequently confirmed this with N-body simulations.       
\end{itemize}
\noindent
The most reasonable  outcome of this debate,
at least in our view and particularly in light of the latter results,
is that observable differences caused by backreaction effects will be of the order of $1\%$.

\subsection{Spatial curvature}

Current constraints on the background spatial curvature, characterized by  $\Omega_{k}$,
within the standard cosmology are
often used to ``demonstrate''  that it is dynamically negligible:  $\Omega_k \sim  5\times 10^{-3} $ (95\% confidence level) \cite{planck2015}. 
However, in standard cosmology the spatial curvature is assumed to be zero (or at least very small and below the
order of other approximations) for the analysis to be valid. Therefore,
strictly speaking, the standard model cannot be used to {\em{predict}}
a small spatial curvature.

In general, $\Omega_{k}$ is {\em{assumed}} to be constrained to be very small primarily based on CMB data. However,
the recently measured temperature and
polarization power spectra of the CMB provides a 99\% confidence level detection
of a negative 
$\Omega_{k} = - 0.044$~ $(+0.018,
-0.015$), which corresponds to a positive spatial curvature \cite{planck2018}.
Direct measurements of the spatial
curvature $\Omega_{k}$  using low-redshift data  such as supernovae, baryon acoustic
oscillations (BAO) and Hubble constant observations
(as opposed to fitting
the FLRW model to the data) do not place tight
constraints on the spatial curvature and allow for a large
range of possible values (but do include spatial flatness).
Low-redshift observations often rely on some  CMB priors \cite{Ratra2017} and, in addition, are sensitive
to the assumptions about the nature of dark energy.
{\footnote{For late Universe observables there is 
significant degeneracy between  $\Omega_{k}$ and dark energy parameters; the standard approach
is to treat 
$\Omega_{k}$ and these parameters as independent quantities, and to marginalize over the
dark energy parameters \cite{XuHuang}.}}

Attempts at a consistent analysis of CMB anisotropy data in the non-flat case suggest a closed model with  $\Omega_{k} \sim 1 \%$ \cite{PR2018,Ryan2019}. Including low redshift data,
$\Omega_k = -0.086 \pm 0.078$ was obtained \cite{PR2018}, which provides weak evidence in favor of a closed spatial geometry (at the level of $1.1\sigma$), with stronger evidence for closed spatial hypersurfaces (at a significantly higher $\sigma$ level) coming from dynamical dark energy models
\cite{Ryan2019} (see also 
\cite{YuRyan}).
The inclusion of CMB lensing
reconstruction and low redshift observations, and especially 
BAO data, gives a model dependent constraint of
$\Omega_{k} = -0.0007 \pm 0.0019$ \cite{planck2018}.


As an illustration, constraints on the phenomenological {\emph{two curvature model}}
(which has a simple parametrized backreaction contribution \cite{CPZ} leading to decoupled spatial curvature
parameters $\Omega_{k_g}, \Omega_{k_d}$  in the metric and the Friedmann equation, respectively, and which reduces to the standard cosmology when  $\Omega_{k_g} = \Omega_{k_d}$),  
were investigated in  \cite{CCCS}.
It was found that the constraints on the two spatial curvature parameters are
significantly weaker than in the standard model, with 
constraints on $\Omega_{k_g}$ an order of magnitude tighter than
those on $\Omega_{k_d}$, and there are tantalizing hints from Bayesian model selection statistics that the data 
favor $\Omega_{k_d} \neq \Omega_{k_g}$ at a high level of confidence.


Observations on recently emerged, {\em present-day} {(large-scale mean)} average negative curvature are weak and not easy to measure \cite{Larena09template}.
Local inhomogeneities and perturbations to the distance-redshift relation at second-order contribute a monopole at the sub-percent level, leading to a shift in the apparent value of the spatial curvature
(as do other GR curvature effects in inhomogeneous spacetimes).
Indeed, in an investigation of how future measurements of  $\Omega_k$ are affected by GR effects, it was shown that constraints on the curvature parameter may be strongly biased if
cosmic magnification is not included in the analysis \cite{DiDio}. 

 Given that current curvature upper limits are at least one order of magnitude away from the level required to probe most of these effects, there is an imperative to continue pushing the curvature parameter, $\Omega_k$, constraints to greater precision (i.e., to about the 0.01\% level). 
These will become increasingly measurable in future surveys such as the Euclid satellite.
In addition, the current curvature parameter estimations are not yet at the cosmic variance limit (beyond which constraints cannot be meaningfully improved due to the cosmic variance of horizon scale perturbations); indeed, the current measurements are more than one order of magnitude away from the limiting threshold \cite{DiDio}.
The prospects for further improving measurements of spatial curvature
are discussed in
\cite{Jimenez}.
Most importantly, we are interested in model independent \cite{Clarksonprl} and explicitly CMB-independent \cite{XuHuang} checks of the cosmic 
flatness. 

However,
currently there is no fully independent constraint with an appropriate
accuracy for a value of
$\Omega_{k}$  of approximately less than  $0.01$ on the cosmic 
flatness from cosmological probes. 
In principle, a small non-zero measurement of 
$\Omega_{k}$ perhaps indicates that the assumptions in the standard model are not met,
thereby motivating models with curvature at the level of a few percent. Such models
are certainly not consistent with simple inflationary models
in which $\Omega_{k}$ is expected to be negligible \cite{Martin19}.
We remark that an observation of non-zero spatial curvature, even at the level of a percent or so, could be the result of backreaction effects
and be a signal of non-trivial averaging effects  \cite{Averaging}. Note that calculations imply a small positive spatial curvature \cite{CPZ} 
(although backreaction estimates have tended to give a negative mean curvature \cite{Larena09template}).

If the geometry of the universe does indeed deviate slightly 
from the standard FLRW geometry (for example, due to the evolution
of cosmic structures), then the spatial curvature
will no longer necessarily be constrained to be constant and any effective spatial flatness may not
be preserved. An investigation of a {\em{small}} emerging spatial
curvature can be undertaken by relativistic cosmological simulations \cite{Computational,Adamek}.
However, such simulations need to include all relativistic corrections
and can  suffer from gauge issues \cite{TB,Adamek}.
In particular, using a fully inhomogeneous, anisotropic cosmological numerical simulation,
it was shown that  \cite{Macpherson}: (i)
On small scales, below the measured homogeneity
scale of the standard cosmology, deviations in cosmological parameters of
6 - 31\% were found 
(in general agreement with LPT and
with deviations depending on an observer's physical location).
(ii) On the approximate homogeneity scale of the Universe
mean cosmological parameters consistent to about 1\%  with
the corresponding standard cosmology were found (although
the parameters can deviate from these mean
values by 4-9\% again depending on the physical location in the simulation
domain). (iii) Above the
homogeneity scale of the Universe,  2 - 3\% variations in mean spatial
curvature and backreaction were found.

As noted above, attempts  to study relativistic models of inhomogeneities  rely upon metric forms that are designed to be ``close to'' the spatially homogeneous and isotropic metric form. However, 
these {\em{can not}} also be used to address the cosmological backreaction problem; backreaction 
can only be present if the  structure--emerging average spatial curvature, and hence the large--scale average of cosmological variables, are allowed to evolve \cite{Bolejko2017}.
A dynamical coupling of matter and geometry on small scales
which allows spatial curvature to vary is a natural feature of
GR. Indeed, the requirement that spatial curvature remains
constant as in
an FLRW model on arbitrarily large scales of cosmological averaging is
not a natural consequence of any principles of GR.
Schemes that suppress average curvature evolution  (e.g., by
employing periodic boundary conditions as in Newtonian models
and neglecting global curvature evolution) can not describe
global backreaction but only ‘cosmic variance’ \cite{BC}.
Moreover, within standard cosmology, spatial fluctuations are conceived to evolve on an assumed
background FLRW geometry, but this
description only makes sense with respect to their spatial average distribution and
its evolution. 
We note that even small fluctuations within averaging schemes are also subject to gauge issues \cite{VPG}. In principle,
large effects are possible from inhomogeneities and  averaging \cite{dust,BC}.

Recently, a relativistic
(Simsilun) simulation based on the approximation of a
`silent universe' was presented \cite{Bolejko18}. The
simulation begins with perturbations around a (flat)
standard model (with
initial conditions set up using the Planck data). The perturbations are allowed to have
non-zero spatial curvature. Initially, the negative curvature
of underdense regions is compensated by the positive
curvature of overdense regions \cite{Roukema19,CPZ}. But once the evolution enters
the non-linear regime, this symmetry is broken and the
mean spatial curvature of the universe slowly drifts from
zero towards negative curvature induced by cosmic voids
(which occupy
more volume than other regions). 
The results of the Simsilun simulation indicate that the
present-day curvature of our Universe is 
$\Omega_k \sim 0.1$, as compared to the spatial flatness of the
early universe.

It should be emphasised 
that the fact that structure formation implies that the present-day Universe (is volume-dominated by voids and) is characterized by on average negative curvature is a subtle issue that follows from the
result that intrinsic curvature does not obey a conservation law \cite{BC,DHB}. 
Indeed, it dispels the naive expectation that on large scales the distribution
of positive spatial curvature for  high-density regions
and negative spatial curvature for the voids, averages out to the
almost or exactly zero spatial curvature assumed. 
    
\newpage

\section{Problems from the quantum realm}

There are a number of very fundamental problems in the quantum regime, culminating in the question of whether there is a single unified theory of quantum gravity (QG).
And, in particular, is this ``theory of everything" string theory? 
Some problems in the quantum realm are relevant for cosmology. For example, do there exist any fundamental
particles that are predicted by QG that have not yet been observed and, if so, what are their properties
and are they of importance in cosmology? In particular,
the detection of the Higgs boson seems to complete
the standard model, but with additional new physics that is needed to protect
the particle mass from quantum corrections (that could increase it by 14
orders of magnitude). 
It is believed that supersymmetry is the most reasonable solution to this
naturalness problem, but the most simple supersymmetric models have
not proved successful and, to date, there is no convincing mechanism to 
break supersymmetry nor to determine the multiple parameters of the supersymmetric theory. In addition,
does a theory of QG lead to a multiverse in cosmology?
And, perhaps most importantly, do theories of QG naturally lead to inflation?

\subsection{The problem of quantum gravity:} 
The standard model of particle physics concerns only 
3 forces: namely, 
electromagnetism and the strong and weak nuclear forces.  
A primary goal of theoretical physics is to derive a theory of QG
in which all 4 forces,
including that of gravitation, are unified within a single field theory.
Up to now, no attempt at such a unification has been successful.
In particular, it is of interest to know whether
QG can be formulated for cosmology and whether there is any extension of quantum mechanics required for QG, and especially quantum cosmology?
Quantum cosmology gives rise to a number questions concerning a possible theory for the initial cosmological state \cite{Hartle}, which include: what laws or principles 
might characterize the initial conditions of the Universe and what are the 
subsequent predictions of the initial conditions for the
Universe on macro-, meso- and
micro-scopic scales?

Let us first briefly discuss two cosmological problems that originate in QG and
have a very distinct mathematical formulation of particular interest here.

\subsubsection{Higher dimensions}

Ordinary spacetime is 4D, but additional
dimensions are possible in, for example, string theory \cite{string}.
At the classical
level, gravity has a much richer structure in higher dimensions than in 4D.
In particular, there has been a lot of work
done on the uniqueness and stability of black holes in arbitrary dimensions \cite{EmparanReall}.
For example, closed trapped surfaces and singularity theorems
in higher dimensions have been discussed \cite{Galloway} and the positive mass theorem 
has been proven in arbitrary dimension \cite{SchoenYau2017}. 
However, the problem of stability in higher dimensions is much more difficult.
Indeed, there is evidence from numerical simulations to indicate that 
there are higher dimensional black holes that are not stable \cite{EmparanReall}.
In addition, the question of
cosmic censorship in higher dimensions is extremely difficult
and is perhaps not even well posed. 
In fact, there is numerical evidence 
that suggests that cosmic censorship does not hold \cite{LehnerPretorius2010} and
that black holes are not necessarily stable to
gravitational perturbations in higher dimensions \cite{GregoryLaflamme}. 
Indeed,
black holes become highly
deformed at very large angular momenta and resemble black branes, and in 
spacetime dimensions greater than six exhibit
an ``ultraspinning instability'' \cite{EmparanMyers}.

Higher dimensional
spacetime manifolds are also considered in a number of cosmological scenarios. 
For example, in the cosmological context all known mathematical results can be
investigated in models with a non-zero cosmological constant.
In addition, theoretical results, such as  the  dynamical stability of higher 
dimensional cosmological models, are of interest. 
In particular, spatially homogeneous cosmologies in higher dimensions, and  
especially extensions of  the  BKL analysis, have been investigated
\cite{Henneaux}.

\subsubsection{AdS/CFT correspondence}

 Anti de Sitter
(AdS) spacetimes are of interest in QG theories formulated in terms of string theory
due to the AdS/CFT (or Maldacena gauge/gravity duality)
correspondence, in
which  string theory on
an asymptotically AdS spacetime is conjectured to be equivalent to a conformally invariant quantum field
theory (CFT) on its boundary \cite{mald,Klebanov}.
This holographic paradigm leads to a number of cosmological questions. 
In particular, the AdS/CFT conjecture
strongly motivates the  dynamical investigation of asymptotically AdS
spacetimes. But, of course, such a spacetime is clearly not this Universe.
In addition,  recently it has been conjectured that AdS spacetimes are unstable
to arbitrarily small perturbations  \cite{PiotrBizon}.

The  global
non-linear stability of AdS has been investigated in spherically
symmetric  massless scalar field models within GR \cite{Bizon}.
Numerical evidence seems to indicate that AdS spacetimes are non-linearly unstable to a ``weakly
turbulent mechanism'' in which an arbitrarily small black hole is formed whose mass is determined
by  the initial  energy. 
Such a
non-linear instability appears to happen for various typical perturbations.
However, there are also many perturbations 
that don't lead to an instability,  which consequently implies the existence of ``islands of 
stability'' \cite{DiasSantos,Martinon}.
It is of great interest to study the non-linear stability of AdS with no assumptions on
symmetry; however, such a study is currently intractable both analytically and numerically.
But the general gravitational case is clearly richer than the  case of spherical symmetry analysed
to date \cite{DiasSantos}. Therefore, it is of great significance to determine 
if the conjectured non-linear instability in AdS spacetime in more general models 
behaves in a similar or a different way to that in
spherically symmetric scalar field collapse  \cite{PiotrBizon}.

\subsection{Singularity resolution}

\subsubsection{Singularity resolution and a quantum singularity theorem} 

The
existence of singularities indicates a failure of GR  when the classical 
spacetime curvature is sufficiently large. This is exactly when QG effects are 
anticipated to be important. Therefore, the problem
of if, and when, QG can extend solutions of classical GR
beyond the singularities is crucial \cite{DeWitt}.
It is, of course, pertinent to determine whether all singularities can be removed in QG.
However, it is certainly not true that all singularities can be resolved within string theory;
for example, 
it is known  that the string in an exact plane wave background  does not 
propagate through the curvature singularity in a well-behaved manner
\cite{Horowitz}.

Gauge/gravity duality, which can be regarded as providing an indirect formulation of string theory \cite{maldecena},
has been utilized to
study singularities in the quantum realm and investigate cosmic censorship 
with asymptotically AdS initial data.
The existence of a quantum version of cosmic censorship was suggested from holographic QG \cite{EngelhardtHorowitz}.
It has been deduced that a large class
of bounces through cosmological singularities are forbidden. Consequently,
although some singularities can indeed be resolved, a novel singularity theorem
is possible. Therefore, it is important to determine 
whether a quantum mechanical generalization
of any of the singularity theorems exists, which would subsequently imply that singularities are inevitable even in quantum settings. 
In particular, it has been shown that a fine-grained generalized second law
of horizon thermodynamics  can
be used to prove the inevitability of singularities \cite{Wall2013}, 
thereby extending the classical singularity  theorem of Penrose  \cite{Penrose1979} to the semi-classical regime. 
It is plausible 
that this result, which was constructed in the context of semiclassical gravity,
will still hold in a complete theory of QG \cite{Wall2013}. Therefore, 
not all singularities can be resolved within QG.

\subsubsection{Cosmological singularity resolution}

Cosmological singularity resolution can be investigated within loop
quantum gravity (LQG) and string theory.
(Black hole singularity resolution was reviewed in \cite{OpenProb}.)
LQG is a non-perturbative canonical
quantization of gravity.
It has been suggested that singularities may be
generically resolved within LQG  as a result of QG effects \cite{Singh14}.
In particular, the classical
big bang singularity is replaced by a symmetric quantum big bounce when
the energy density is of the order of the Planck  density, which
occurs without any violation of
the energy conditions or fine tuning. The resulting quantum big bounce
connects the currently expanding universe to a pre-bounce contracting
classical universe.

The application of LQG in the context of cosmology
is referred to as loop quantum cosmology (LQC). In LQC the
infinite  degrees of freedom reduce to a finite number due to spatial homogeneity. 
A variety of spatially homogeneous cosmologies have been investigated
\cite{AshtekarandSingh}. In particular, solutions to the effective equations 
for the general class of Bianchi type IX cosmological
spacetimes has been investigated within LQC computationally, wherein the big bang singularity was shown to be resolved \cite{Ewing2010}. 
The reduction of symmetries within LQC involves a very considerable simplification, and consequently 
crucial aspects of the dynamics may be neglected.
However,  partly due
to evidence supporting
the BKL conjecture, 
it is believed that the singularity resolution in spatially homogeneous cosmologies does
capture important features of singularity resolution in more general spatially  inhomogeneous 
cosmological models
\cite{AshtekarandSingh,Brizuela}.
There are ongoing attempts to include spatial
inhomogeneities in the analysis \cite{Tarrio}.

Various  singularities 
have been investigated within standard LQC.
It has been conjectured that all curvature singularities which result in
geodesic incompleteness are so-called strong singularities (such as the big bang in GR). In recent years a number of other types of cosmological singularities  have
been obtained, which include the big rip and the  big freeze, and sudden and 
generalized sudden singularities. Of these, the big rip and big freeze are strong
singularities within GR, whereas sudden and generalized sudden singularities are weak singularities.
Using a phenomenological matter model in GR, 
it has been established that strong singularities are, in  general, resolved in
LQC, whereas  quantum geometry does not usually affect
weak singularities \cite{SainiSingh}.
A comprehensive investigation of the resolution of
a variety of singularities within modified LQC models, in which the bounce can be asymmetric
and the  bounce density can be affected,
was performed using an effective spacetime description
and compared with the analysis in standard LQC  \cite{SainiSingh}.

\subsection{Quantum gravity and inflation}

Although some of the alternatives to inflation alluded to earlier are suggested by ideas motivated by QG,
it is also of interest to determine whether inflation occurs naturally within QG.
For example, it appears to be difficult to get inflation within string theory  \cite{deS}.
In particular, so-called swampland criteria constrain inflationary models and
there are no-go theorems for
the existence of de-Sitter vacua in critical string theory.
The fact that exact de Sitter solutions with a positive cosmological constant cannot 
describe the late-time behaviour of the Universe \cite{deS} is often interpreted as ``bad news'' for string theory.

The observations of Planck 2018
(of the almost scale-invariant and Gaussian primordial curvature
perturbations) \cite{planck2018}
are compatible with the predictions
of simple single
scalar field inflation models
with a canonical kinetic term and an appropriately
flat self-interaction potential  minimally coupled to gravity. However, despite 
the success of the single-field slow-roll inflation model, it is  not straightforward to embed
such a model within a
fundamental theory \cite{baumann}.

However,
the so-called $\alpha$-attractor models and, in particular, the  KKLMMT model  \cite{KKLMMT}, have been actively studied. 
The most attractive theoretical properties of these models is
their conformal symmetry and their successful embedding
into supergravity via hyperbolic geometry.
The KKLMMT
model is often acknowledged as the first to discuss
the origin of D-brane inflation 
within string theory \cite{KKLMMT},
and provides the motivation for  more general string inspired
cosmological models. 
These models predict
values for the spectral index and the tensor-to-scalar ratio which match observational data well. 
Thus, phenomenological D-brane inflation has attained renewed
importance, independent of its string theory origin, since Planck 2018 \cite{planck2018}.  
Indeed, it has been shown
\cite{Kall} that 
further phenomenological models of D-brane inflation
can be derived within the string theory approach (see also \cite{baumann}).
Because scalar fields (such as, for example, moduli fields) occur
ubiquitously in fundamental theories such as supergravity and
string/M theory, multi-field generalizations of the $\alpha$-attractor models
have also been considered \cite{double}.

A number of inflationary  cosmologies have been suggested within the context of
string/M-theory \cite{KKLMMT,deS}. However, very few models exist that can be embedded within LQC \cite{Kiefer}. In particular,
there are a number of approaches to QG which include bouncing regimes.
In resolving the initial singularity, it is of interest to determine whether slow-roll inflation
is subsequently allowed (or is even  natural). 
Inflation within the context of LQC, 
and how the bounce affects the  evolution of the inflaton (as compared to the normal scenario with no bounce), was investigated in \cite{Louko}.
The evolution of the inflaton from the
initial bounce was studied analytically for a number of important potentials
in the case that the inflaton is 
taken to be the same scalar field that gives rise to the LQC bounce.
It was found \cite{Louko} that LQC, or any bouncing model in which the 
total energy density   of the inflaton field is bounded at the transition, does
provide a viable description of the pre-inflationary epoch and the subsequent smooth evolution to the standard inflationary era.
The results were particularly encouraging in  that the bounds obtained theoretically (on the critical bounce value for the inflaton field
in order for there to subsequently be an appropriate slow roll inflationary regime) match (where appropriate)  the known results
from the numerical dynamics of the fully non-linear LQC.

\subsubsection{String inflation}

Cosmological inflation and its realization within QG and, in particular, 
in string theory, was reviewed in \cite{baumann}.
Examples of string inflation include  brane and axion inflation. There are also string inspired effective field theories.
Since string theory is  considerably more constrained,
some effective field theories that are apparently consistent at low energies do
not, in fact, admit ultraviolet QG completions 
(leading to improved predictivity). However,
there are indications that it might not be possible
to embed simple inflationary models in string theory \cite{deS,baumann}. One problem is that in order to obtain a period of slow-roll inflation from simple scalar field
potentials, field values in excess of the Planck mass are required. 
But for sufficiently  
large values of the fields string effects on the shape of the potential 
must be included, which  tend to destroy its required
flatness except perhaps in the case of special  field symmetries  \cite{beyond} (however, 
even then string theoretical arguments such as the so-called ``Weak Gravity Conjecture'' 
\cite{Arkani-Hamed} can lead to
the effective
field theory analysis being invalidated).

In addition,
generating effective theories from  string theory can also lead
to different ideas as to what a natural (or
a minimal) inflationary model might be. 
Indeed, a comprehensive understanding of
naturalness within string theory is elusive. However,
a general feature  of all stringy constructions 
is the existence of a number of light scalar fields, so while
multiple `unnecessary' fields might be considered non-minimal in many field theory
models, they are ubiquitous  within string theory.
Time-dependent solutions with string scale curvatures 
are crucial for any further comprehension,
especially if we hope
to progress from the paradigm of an effective theory for the massless modes.

To date, it is fair to say that there have  not been any convincing
realizations of inflation in the context of superstring theory. 
Making predictions in string theory is made exceedingly difficult by the {\em{landscape}}
problem  that string theory has an enormous
number of vacua. 
Despite the fact that dynamics within the
landscape is not well understood, it appears that
false vacuum eternal inflation
is an unavoidable consequence. 
In addition,
all 4D de Sitter vacua in supersymmetric string theories are
metastable, 
since 10D supersymmetric  Minkowski spacetime has zero energy, but de Sitter spacetime has
positive vacuum energy. 
In particular, there are well-known no-go theorems for
the existence of stable de-Sitter vacua in critical string theory \cite{deS}.
This is a real problem for inflation should
string theory be the final theory of QG.

The so-called string {\em{swampland}} criteria constrain inflationary models  \cite{Denef}.
In addition, the
second of the swampland conjectures implies, as noted above,
that exact de Sitter solutions with a positive cosmological constant cannot describe the
fate of the Universe at late times  within string theory \cite{deS}.
Dynamical dark energy scalar field  models
must also satisfy
particular criteria so as to avoid the swampland.
The observational implications of such string-theory criteria
on quintessence models and the
accompanying constraints
on the  dark energy were studied in \cite{Heisenberg}.
However, since string theory does not naturally lead to scalar fields
with an appropriate energy scale 
to be a reasonable candidate for quintessence,
novel physics from
string theory must be introduced to explain
dark energy.
In some very special models it is possible to
characterize the Planck-suppressed corrections to the  string theory  inflatonary action, 
leading
to the first indications for inflation within string theory \cite{baumann}. 
But many critical challenges still remain. 
Indeed, the `simple' cosmological observations (of the almost scale-invariant and Gaussian primordial curvature perturbations measured by Planck) to date
are often interpreted as an argument against complex models of inflation in string theory (however, see
\cite{baumann}).

\newpage

\section{Concluding remarks}

We have reviewed recent developments and described a number of open questions 
in the field of theoretical
cosmology. We described
the concordance cosmological model and the standard paradigms of modern cosmology, and then 
discussed a number of fundamental issues and open theoretical questions,
emphasizing the various assumptions made and identifying which results are independent of these assumptions.
Indeed, standard cosmology  contains a number of philosophical assumptions that are not 
always scientific, including 
the assumption of spatial homogeneity and isotropy at large scales
outside our particle horizon.
Perhaps a more tangible fundamental issue concerns  
the measure problem and the issue of initial conditions in inflation.
Many of the fundamental  problems arise due to the inhomogeneities in the Universe. However, this is also one of the great strengths of present day cosmology: our models predict what structure will occur, and consequently the astounding development of observational projects determining in great detail the characteristics of such structure that serve to give strong limits on cosmological parameters.

Cosmology is not only a mathematical endeavour, but it is a testable scientific theory 
due to its ability to produce observational predictions. In recent times there has been a plethora of such detailed tests, leading to the so-called era of precision 
cosmology. Perhaps fundamental questions  are less relevant for current working cosmologists,
who are more concerned with physical cosmology and data and statistical analysis. 
But as the modern emphasis changes to more physical and observational issues,
theoretical cosmology is 
still important and fundamental questions persist.
In some sense, we hope to record here the  state of the art as it 
now exists.  

A qualitative analysis of the properties of cosmological models and 
the problems of the stability of cosmological solutions and of
singularities is important in mathematical cosmology. A number of open problems in theoretical cosmology involve 
the nature of the origin and 
details of cosmic inflation, and its relation to fundamental physics. Perhaps the most urgent open problems
of theoretical cosmology include the early
and late time accelerated expansion of the universe
and the role of the cosmological constant $\Lambda$. 
As we have emphasized,  computational cosmology is becoming an increasingly important
tool in the investigation of theoretical and physical cosmology.

We then reviewed a number of  open problems in
physical cosmology, with particular focus on perturbation theory (and gauge issues)
and the formation and distribution of large scale structure in
the Universe at present times (and especially in the non-linear regime).
Backreaction is still an important issue, although perhaps the more formal mathematical averaging
problem is currently more relevant.
Finally, gravitational wave
astronomy will potentially play an increasingly important role within 
cosmology. Indeed, there is a 
robust prediction within inflation for a gravitational wave
induced CMB polarization signal.

We also discussed 
cosmological problems in quantum gravity, including the 
possible resolution of cosmological  singularities
and the crucial issue of the role of inflation within
quantum gravity. 

Finally we have emphasized that, given the uniqueness of the Universe and the limitations on the domain we can explore by any conceivable observations, it is key to carry out all possible consistency tests of our models. For example, the first and foremost is the age of the universe: is the Universe older than its stellar and galactic content? If not, cosmology is in deep trouble. Fortunately this 
consistency test
seems to be satisfied at present (thanks to the cosmological constant). Another 
consistency test is that all number counts must display a dipole aligned with the CMB dipole; this is presently being contested.

\newpage

\section*{Acknowledgements}  
We would like to acknowledge Bernard Carr for his helpful comments, and to thank Timothy Clifton and Julian Adamek for fruitful discussions. 
Financial support was provided by  NSERC of Canada (AAC) and 
NRF of South Africa (GFRE).



\begin{thebibliography}{99}

\baselineskip 12pt

\bibitem{Ellis1971} G. F. R. Ellis, {\em Relativistic
Cosmology} in {\em General Relativity and Cosmology}, Proc. Int.
School of Physics `Enrico Fermi' (Varenna) Course XLVII pp104-179,
ed. R. K. Sachs (Academic Press, 1971).


\bibitem{krameretal}
H. Stephani, D. Kramer, M. MacCallum, C. Hoenselaers and E. Herlt, ``Exact Solutions of Einstein’s
Field Equations'' (Cambridge University Press, Cambridge, second ed., 2003).


\bibitem{fit}
G. F. R.~Ellis,
``Relativistic cosmology: Its nature, aims and problems'',
in  General
Relativity and Gravitation, pp.~215--288 eds. B.~Bertotti, F.~de Felice and A.~Pascolini (Reidel, Dordrecht, 1984);
G.F.R.~Ellis and W.~Stoeger,
Class.  Quant. Grav.  {\bf4} 1697 (1987);
S. Bildhauer and
T. Futamase, Gen. Rel. Grav. {\bf{23}} 1251 (1991; T. Futamase, Phys. Rev. D {\bf{53}}
681 (1993); J. P. Boersma, Phys. Rev. D {\bf{57}}  798 (1998).

\bibitem{Bertotti}
B. Bertotti, Proc. Roy. Soc. London A {\bf{294}}
195 (1966). 

\bibitem{OpenProb} A. Coley,  Phys. Scr. {\bf{92}} 093003 (2017).

\bibitem{Hilbert}
D. Hilbert, 
Bull. Amer. Math. Soc. {\bf{8}} 437 (1902)  
(see also, in the original German, Gottinger Nachrichten {\bf{1}} 253 (1900)
\& Archiv. Math. Phys.  {\bf{1}}  44 \& 213 (1901)).



\bibitem{simon} B. Simon, ``Fifteen Problems in Mathematical Physics,
Perspectives in Mathematics'',
Anniversary of Oberwolfach 
at Birkhiiuser Verlag, Basel (1984).




\bibitem{gonitsora}
http://gonitsora.com/fivegreatunsolvedproblemsintheoreticalphysics.



\bibitem{MathGR} A. Coley, Gen. Rel. Grav. {\bf{51}} 78 (2019) [arXiv:1807.08628].


\bibitem{SenovillaGarfinkle}
J. M. M. Senovilla and D. Garfinkle, 
Class. Quant. Grav. {\bf{32}}  124008 (2015)
[arXiv:1410.5226]. 

\bibitem{Dotti}
G. Dotti, Phys. Rev. Lett. {\bf{112}}  191101 (2014) \&
Class. Quant. Grav. {\bf{33}}    205005 (2016).


\bibitem{Ellis2014} G. F. R. Ellis, Studies in History and
Philosophy of Modern Physics, {\bf{46}} 5 (2014).

\bibitem{Butterfield} J. Butterfield, Studies in History and
Philosophy of Modern Physics, {\bf{46}} 57 (2014).

\bibitem{Sahlena} M. Sahlena, ``On Probability and Cosmology: Inference Beyond
Data?" [arXiv:1812.04149];
Slightly expanded version of contribution to the book ‘The Philosophy of Cosmology’, eds. K.  Chamcham, J. Silk, J. D. Barrow and S. Saunders
(Cambridge University Press, 2017).

\bibitem{measure}
A. Linde and M. Noorbala,  J. Cosmol. Astropart. Phys. {\bf{9}} 008 (2010); K. D.
Olum, Phys. Rev. D {\bf{86}} 063509 (2012).

\bibitem{SilkLimits} J. Silk,
Found. Phys. {\bf{48}} 1305 (2018).


\bibitem{Weinberg1987} S. Weinberg, 
Phys. Rev. Lett. {\bf{59}} 2607 (1987).


\bibitem{EllisSilk} G.
Ellis and J. Silk, Nature {\bf{516}} 321 (2014).

\bibitem{Maartens_Homog}
R. Maartens,  Phil. Trans. R. Soc. A  {\bf{369}} 5115 (2011).

\bibitem{EllMaaMac} G. F. Ellis, R. Maartens and  M. A. MacCallum, {\em{Relativistic cosmology}} (Cambridge University Press, 2012).


\bibitem{sl09}
F.~Sylos Labini, N.L.~Vasilyev, L.~Pietronero and Y. V.~Baryshev,
Europhys.\ Lett.\ {\bf 86}  49001 (2009) [arXiv:0805.1132]; see also [arXiv:1512.03313].



\bibitem{dust}
D. L.~Wiltshire,
Class.  Quant. Grav.  {\bf28} 164006  (2011) [arXiv:1106.1693].


\bibitem{DHB}
C. Desgrange, A. Heinesen and T. Buchert,
``Dynamical spatial curvature as a fit to type Ia supernovae''
IJMPD [arXiv:1902.07915].



\bibitem{BC}
T.~Buchert and M.~Carfora,
Phys. Rev. Letts. {\bf 90} 031101  (2003) [gr-qc/0210045];
D.~L.~Wiltshire,
New J. Phys. {\bf{9}} 377  (2007) [arXiv:gr-qc/0702082];
T. Buchert, ``Is Dark Energy Simulated by Structure Formation in the Universe?'' [arXiv:1810.09188].



\bibitem{Adamek} J. Adamek, D. Daverio, R. Durrer and M. Kunz,  J. Cosmol. Astropart. Phys. {\bf{2016(07)}} 053 (2016).
J. Adamek, D. Daverio, R. Durrer and M. Kunz,  Nature
Physics {\bf{12}} 346 (2016) [arXiv:1509.01699]; J. Adamek, C. Clarkson, D. Daverio, R. Durrer and M. Kunz,
Class. Quant. Grav.
{\bf{36}} 014001 (2019)
[arXiv:1706.09309]; J. Adamek, D. Daverio, R. Durrer and M. Kunz,  J. Cosmol. Astropart. Phys.  {\bf{2016}} 053
(2016).


\bibitem{Macpherson}
H. Macpherson, D. J. Price and P. D. Lasky, 
``Einstein's Universe: cosmological structure formation in numerical relativity''
[arXiv:1807.01711]; H. Macpherson, P. D. Lasky and D. J. Price,  ``The trouble with Hubble''
[arXiv:1807.01714].



\bibitem{Cornish}
N. J. Cornish, D. N. Spergel, and G. D. Starkman, Class. Quant. Grav. {\bf{15}} 2657 (1998).

\bibitem{small}
R. Aurich, H. S. Janzer, S. Lustig and F. Steiner, 
Class. Quant. Grav. {\bf{25}} 125006 (2008); see also
G. F. R. Ellis and G. Schreiber, Phys. Lett. A {\bf{115}} 97 (1986).



\bibitem{Hartle} J. B. Hartle,
``The Impact of Cosmology on Quantum Mechanics''
[arXiv:1901.03933];
J. B. Hartle, ``Quantum Cosmology: Problems for the 21st Century''
in Proceedings of the 11th Nishinomiya-Yukawa Symposium, ed by K. Kikkawa {\em{et al.}}, World Scientific Singapore, 1998. [arXiv:gr-qc/9701022]. 


\bibitem{Carrmultiverse}
B. J. Carr (editor),
{\em{Universe or Multiverse}}
(Cambridge University Press, 2007); also see
B. J. Carr and G. F. R. Ellis, Astron. Geophys.
{\bf{49}} 2 (2008).


\bibitem{HawEll73} S. W. Hawking and  G. F. R. Ellis, \textit{The large scale structure of spacetime} (Cambridge Univ. Press, Cambridge, 1973).


\bibitem{LL} L. D. Landau and E. M. Lifshitz, {\em Fluid Mechanics}
(Pergammon Press, Oxford, 1987).

\bibitem{Deb} F. Debbasch, Eur. Phys. J. {\bf B37} 257 (2004) \&
{\bf B43} 143 (2005).




\bibitem{CHP}
A. Coley, S. Hervik and N. Pelavas, Class. Quant. Grav. {\bf 26} 025013 (2009)
[arXiv:0904.4877]; A. Coley, S. Hervik and N. Pelavas, 
Class. Quant. Grav. 
{\bf{27}} 102001 (2010)
[arXiv1003.2373]; see also 
A. Coley and S. Hervik, 
Gen. Rel. Grav.
{\bf{43}} 2199 (2011).




\bibitem{Ishak}
M. Ishak, ``Testing general relativity in cosmology''
[arXiv:1806.10122].

\bibitem{Clifton_Ferreira}
T. Clifton, P. G. Ferreira, A. Padilla and C. Skordis ``Modified gravity and cosmology''
Physics reports \textbf{513} 1-189 (2012).



\bibitem{Rendall2002} A. Rendall,  
Living Rev. Rel. {\bf{5}} 6 (2002)
[arxiv:gr-qc/0203012].                  


                                                    

\bibitem{LARS99}
L. Andersson, ``The global existence problem in general relativity,
The Einstein equations and the large scale behavior of gravitational fields'', pp. 71--120 (Birkhäuser, Basel,  2004) [arxiv/gr-qc/9911032].


\bibitem{CB69} Y. Choquet-Bruhat and R. Geroch, 
Comm. Math. Phys. {\bf{14}} 329 (1969).










\bibitem{Narita} M. Narita,
Class.
Quant. Grav. {\bf{19}}  6279 (2002)
[arXiv:gr-qc/0210088].


\bibitem{Penrose1979}
 R. Penrose, ``Singularities and time asymmetry'', in ``General Relativity: an Einstein Centenary
Survey'', eds.  S. W. Hawking and W. Israel (Cambridge University Press, Cambridge, 1979).




\bibitem{Penrose65}
R. Penrose,
Phys. Rev. Lett. {\bf{14}} 57
(1965).


\bibitem{HawEll68} S. W. Hawking and  G. F. R. Ellis,  Ap. J. {\bf{152}} 25 (1968);


\bibitem{Hawking1966} S. W.
Hawking,  
Proc. Roy. Soc. London
{\bf{A294}}  511 (1966); {\em{ibid.}}, 
{\bf{A295}}  490 (1966);  {\em{ibid.}},   {\bf{ A300}} 
187 (1967).


\bibitem{PenroseHawking} R. Penrose and S. W. Hawking,
Proc. Roy. Soc. Lond. A {\bf{314}} 529 (1970).



\bibitem{Senovilla2012}
J. M. M. Senovilla, ``Singularity theorems in general relativity: achievements and open questions'',
Chapter 15 of Einstein and the Changing Worldviews of Physics, eds. C. Lehner, J. Renn and M.
Schemmel, Einstein Studies 12 (Birkhauser, 2012)



\bibitem{Christodoulou2009}
D. Christodoulou, ``The formation of black holes in general relativity'' (Monographs in Mathematics,
European Mathematical Soc. Publishing House, Helsinki, 2009).



\bibitem{Klainerman2014}
S. Klainerman, J. Luk and I. Rodnianski,
Invent. Math. {\bf{198}} 1 (2014).


\bibitem{Klainerman2012}
S. Klainerman and I. Rodnianski, 
Acta Math. {\bf{208}} 211 (2012);
J. Luk and I. Rodnianski, ``Nonlinear interactions of impulsive gravitational waves for the vacuum Einstein
equations'',  Cambridge J. Math. [arXiv:1301.1072]; M. Dafermos, 
Astrisque {\bf{123}} 352 (2013).




\bibitem{Brandenberger}
R. Brandenberger and P. Peter,  Found. Phys. {\bf{47}} 797 (2017) [arXiv:1603.05834].




\bibitem{Khoury} J. Khoury, B. A. Ovrut, P. J. Steinhardt and N. Turok, Phys. Rev. D {\bf{64}} 123522 (2001) [hep-th/0103239];
J. Khoury, B. A. Ovrut, N. Seiberg, P. J. Steinhardt, and N. Turok, Phys. Rev. D {\bf{65}} 086007
(2002) [hep-th/0108187];
P. L. McFadden, N. Turok, and P. J. Steinhardt, Phys. Rev. D {\bf{76}} 104038 (2007)
[hep-th/0512123]; J.-L. Lehners and N. Turok, Phys .Rev. D {\bf{77}} 023516 (2008) [hep-th/0708.0743].



\bibitem{Penrosebooks}
R. Penrose,
``The Emperor's New Mind: Concerning Computers, Minds, and The Laws of Physics'' (Oxford University Press, 1989). 




\bibitem{BrandenbergerVafa} R. H. Brandenberger and C. Vafa, Nucl. Phys. B {\bf{316}} 391 (1989);
A. Nayeri, R. H. Brandenberger, and C. Vafa, Phys. Rev. Lett. {\bf{97}} 021302 (2006);
R. H. Brandenberger,  Class. Quant. Grav. {\bf{28}} 204005 (2011)
[arXiv:1105.3247].


\bibitem{Bruni_bounce} C.
Ganguly and  M. Bruni,  ``Quasi-isotropic cycles and non-singular bounces in a Mixmaster cosmology'' [arXiv:1902.06356].


\bibitem{Cai}
Y. F. Cai, R. Brandenberger and P. Peter,  Class. Quant. Grav. {\bf{30}} 075019 (2013) [arXiv:1301.4703].



\bibitem{Penrosebooks1}
R. Penrose,
 ``Fashion, faith, and fantasy in the new physics of the universe'' (Princeton University Press, 2016).

           


\bibitem{EKp}
D. Garfinkle, W. C. Lim, F. Pretorius and P. J. Steinhardt,  Phys. Rev. D {\bf{78}} 083537 (2008);
B. Xue, D. Garfinkle, F. Pretorius and P. J. Steinhardt, Phys. Rev. D {\bf{88}} 083509 (2013).




\bibitem{Ijjas2}
A. Ijjas, P. J. Steinhardt, and A. Loeb, Phys. Rev. D {\bf{89}} 023525 (2014);
A. Ijjas, J.-L. Lehners, and P. J. Steinhardt, Phys. Rev. D {\bf{89}} 123520 (2014);
A. Ijjas and P. J. Steinhardt,``Bouncing Cosmology made simple''
[arXiv:1803.01961]. 

\bibitem{Turok}
N. Turok, M. Perry and P. J. Steinhardt, Phys. Rev. D {\bf{70}} 106004 (2004).




\bibitem{Bojowald}
M. Bojowald, Phys. Rev. Lett. {\bf{95}} 091302 (2005).

\bibitem{Ashtekar}
A. Ashtekar, T. Pawlowski and P. Singh, 
Phys. Rev. D {\bf{74}} 084003 (2006); P.
Diener, B. Gupt  and P. Singh, 
Class. Quant. Grav. {\bf{31}} 105015 (2014).


\bibitem{bounce}
G. F. R. Ellis, E.  Platts, D. Sloan and  A. Weltman,   J. Cosmol. Astropart. Phys.  {\bf{2016(04)}} 026 (2016).
 

\bibitem{Olinyk}
H. Friedrich,
J. Diff. Geom. {\bf{34}} 275 (1991);
R. A. Bartnik, M. Fisher and T. A. Olinyk,
J. Math. Phys. {\bf{51}} 032504 (2010) 
[arXiv:0907.3975].




\bibitem{Bizonwavemaps} P.
Bizon, 
Comm. Math. Phys. {\bf{215}} 45 (2000); P.
Bizon, T. Chmaj and  Z. Tabor, 
Nonlinearity {\bf{14}} 1041 (2001).


\bibitem{AnderssonG}
L. Andersson, N. Gudapati and J. Szeftel, Ann. PDE {\bf{3}} 13 (2017)
[arXiv:1501.00616];
J. Sterbenz and D. Tataru,
Comm. Math. Phys. {\bf{298}}    231 (2009) 
[arXiv:0907.3148];  P. Bizon and P. Biernat, Comm. Math. Phys. {\bf{333}} 1443 (2015); 
P. Bizon, Acta Physica Polonica  B {\bf{33}} 1893 (2002);
H. Andreasson, 
 Living Rev. Rel. {\bf{14 }}  4 (2011) 
[arXiv:1106.1367].



\bibitem{differentiability} D.
Eardley and V. Moncrief,  
Comm. Math. Phys. {\bf{83}} 171 (1982)
\& Comm. Math. Phys. {\bf{83}} 
193 (1982); S.
Klainerman and M. Machedon, 
Ann. Math. {\bf{142}} 39 (1995);
P. T. Chrusciel and J. Shatah, 
Asian
J. Math. {\bf{1}} 530 (1997); S. Kichenassamy and A. D. Rendall, 
Class. Quant. Grav. {\bf{15}}  1339 (1998).


\bibitem{Bunting}
W. Israel,
Comm. Math.
Phys. {\bf{8}}  245 (1968);
G. Bunting and A. K. M. Masood-ul-Alam, 
Gen. Rel. Grav. {\bf{19}}
147 (1987).

\bibitem{Carter}
 B. Carter, in Black Holes, 1972 Les Houches Lectures, eds. B. DeWitt
and C. DeWitt (Gordon and Breach, NY, 1973); B. Carter, 
Comm. Math. Phys. {\bf{99}}
563 (1985);
D. C. Robinson, 
Phys. Rev. Lett. {\bf{34}} 905 (1975).



\bibitem{Newman65} E. Newman, J. Math. Phys. {\bf{6}} 918 (1965);
P. Mazur,
J. Phys. A {\bf{15}} 3173 (1982).




\bibitem{ChristodoulouKlainerman90} D.
Christodoulou, and  S. Klainerman,
Commun. Pure Appl. Math. {\bf{43}}
137 (1990).




\bibitem{Christodoulou93}
D. Christodoulou and S. Klainerman, ``The global nonlinear stability of the Minkowski space''
(Princeton mathematical series, 41, Princeton University Press, 1993).

         
 

\bibitem{DHR}
M. Dafermos, G. Holzegel and I. Rodnianski, ``The linear stability of the Schwarzschild solution to gravitational perturbations'',
146 pages (2016) [arXiv:1601.06467]; G. Holzegel,
Class. Quant. Grav. {\bf{33}}  205001 (2016).



\bibitem{Heusler}  
S. Chandrasekhar, ``Mathematical Theory of Black Holes'' (Oxford University Press, 1983);
M. Heusler,
Living Rev. Rel. {\bf{1}} 6 (1998);
http://www.livingreviews.org/Articles/Volume1/1998-6heusler.



\bibitem{KlainermanSzeftel}
 S. Klainerman and J. Szeftel, 
``Global Nonlinear Stability of Schwarzschild Spacetime under Polarized Perturbations'',
425 pages [arXiv:1711.07597].

\bibitem{Shlapentokh}
M. Dafermos and I. Rodnianski,
``Lectures on black holes and linear
waves'', Clay Mathematics
Proceedings {\bf{17}} 97 (2008) [arXiv:0811.0354];
M. Dafermos, G. Holzegel and I. Rodnianski,
 ``Boundedness and decay for  the  Teukolsky equation on Kerr spacetimes I'' 
[arxiv/1711.07944].





\bibitem{Zilho}
M. Zilhao, V. Cardoso, C. Herdeiro, L. Lehner and U. Sperhake, 
Phys. Rev. D {\bf{90}} 124088  (2014) [arXiv:1410.0694].



\bibitem{Computational}
A.~Coley, L. Lehner, F. Pretorius and D. ~Wiltshire,
``Computational Issues in Mathematical Cosmology'' (2017);
http://cms.iopscience.iop.org/alfresco/d/d/workspace/SpacesStore/
83f10d6e-0b33-11e7-9a47-19ee90157113/Overview-CC.pdf


\bibitem{Bentivegna}
E. Bentivegna and
M. Bruni,  Phys.  Rev.  Lett.  {\bf{ 116}} 251302 (2016) [arXiv:1511.05124]; E. Bentivegna, Phys.
Rev. D {\bf{95}} 044046 (2017) [arXiv:1610.05198].

\bibitem{Giblin}
J. T.  Giblin, J. B.
Mertens and G. D. Starkman,  Phys.  Rev.  Lett.  {\bf{ 116}} 251301 (2016),
{\em{ibid.}} Phys.  Rev.  D {\bf{93}} 124059 (2016) [arXiv:1511.01105],
{\em{ibid.}} Class. Quant. Grav.
{\bf{34}} 214001 (2017)
[arXiv:1704.04307].


\bibitem{Adamek18}
J. Adamek, C. Clarkson, L. Coates, R. Durrer and M. Kunz,
``Bias and scatter in the Hubble diagram from cosmological large-scale structure''
[arXiv:1812.04336].




\bibitem{Lemait27} G. Lema\^{i}tre, Ann.  Soci\'{e}t\'{e} Sci. de Bruxelles \textbf{47} 49 (1927).


\bibitem{Disney}
 M. J. Disney, Nature {\bf{263}} 573 (1976).



\bibitem{tension} 
T. Buchert, A. A. Coley, H. Kleinert, B. F. Roukema and D. L. Wiltshire, Int. J. Mod. Phys.
D {\bf{25}} 1630007  (2016) [arXiv:1512.03313].




\bibitem{planck2018} 
Y. Akrami {\em{et al.}},  \emph{Planck 2018 results. I. Overview} [arxiv:1807.06205];
N.~Aghanim {\em{et al.}} \emph{ VI. Cosmological parameters}, [arxiv:1807.06209];
Y. Akrami {\em{et al.}},  \emph{X. Constraints on inflation} [arxiv:1807.06211].


\bibitem{Finelli}
F. Finelli, J. Garca-Bellido, A. Kovcs, F. Paci and
I. Szapudi, Mon. Not. Roy. Astron. Soc. {\bf{455}} 1246 (2016)
[arXiv:1405.1555];
A. Kovacs and J. Garca-Bellido, Mon. Not. Roy. Astron.
Soc.  {\bf{462}} 1882 (2016) [arXiv:1511.09008].


\bibitem{lithum}
Particle Data Group, Rev. Part. Phys. Chin. Phys. C {\bf{40}} 100001 (2016).


\bibitem{EllisBaldwin}
G. F. R. Ellis  and J. E. Baldwin, Mon. Not. R. Astron. Soc. {\bf{206}} 377 (1984).


\bibitem{WiltshireHubble}
D. L. Wiltshire, P. R. Smale, T. Mattsson, T. and R. Watkins,  Phys. Rev. D
{\bf{88}} 083529 (2013); J. H. McKay and D. L. Wiltshire,   Mon. Not. R. Astron. Soc., {\bf{457}} 3285 (2016) (Err. ibid. {\bf{463}} 3113); see also D. L. Wiltshire, ``Comment on "Hubble flow variations as a test for inhomogeneous cosmology'' [arXiv:1812.01586].

\bibitem{Kraljic} D.
Kraljic and S. Sarkar,   J. Cosmol. Astropart. Phys. {\bf{1610}}  016 (2016).

\bibitem{Maartens_dipole} R.
Maartens, C. Clarkson  and  S. Chen,   J. Cosmol. Astropart. Phys. \textbf{1801} 013 (2018).


\bibitem{Kashlinsky}
A. Kashlinsky, F. Atrio-Barandela and H. Ebeling,  Astrophys. J. {\bf{732}} 1 (2011);
A. Kashlinsky, F. Atrio-Barandela, H. Ebeling, A. Edge, and D. Kocevski, Ap. J. Lett. {\bf{ 712}} L81(2010) [arXiv:0910.4958];
H. A. Feldman, R. Watkin, and M. J. Hudson, Mon. Not. R. Astron. Soc. {\bf{407}}  2328 (2010) [arXiv:0911.5516].

\bibitem{Sarkar}
J. Colin, R. Mohayaee, S. Sarkar and A. Shafeloo, 
Mon. Not. R. Astron. Soc. {\bf{ 414}} 264 (2011) [arXiv:1011.6292];
A. Green, AAO Observer Number 122 (August 2012) [arXiv:1210.0625].


\bibitem{R16}
A.~G. Riess et~al., Ap. J.
{\bf{826}} 56  (2016) [arxiv:1604.01424].
  


\bibitem{tensionRiess} A. G. Riess, S. Casertano, W. Yuan, L. M. Macri and D. Scolnic,
``Large Magellanic Cloud Cepheid Standards Provide a 1\% Foundation for the Determination of the Hubble Constant and Stronger Evidence for Physics Beyond $\Lambda$CDM'' [arXiv:1903.07603].




\bibitem{planck2015}
P.~A.~R. Ade {\em{et~al.}}, {\em{Planck 2015
 results. XIII. cosmological parameters}}, Astron. Astrophys. {\bf {594}} A13 (2016) [arxiv:1502.01589].
  


\bibitem{beyond} R. H. Brandenberger,
``Beyond Standard Inflationary Cosmology''
[arXiv:1809.04926]
(modified version of a contribution to "Beyond Spacetime" eds. N. Huggett, K. Matsubara and C. Wuethrich (Cambridge Univ. Press, Cambridge, 2018)).


\bibitem{Notari}
A. Notari and A. Riotto,  Nucl. Phys. B {\bf{644}} 371 (2002)
[hep-th/0205019];
F. Finelli,  Phys. Lett. B {\bf{545}} 1 (2002) [hep-th/0206112];
F. Di Marco, F. Finelli and R. Brandenberger, Phys. Rev. D {\bf{67}} 063512 (2003) [astro-ph/0211276];
J. L. Lehners, P. McFadden, N. Turok and P. J. Steinhardt, Phys. Rev. D {\bf{76}} 103501 (2007) [hep-th0702153];
E. I. Buchbinder, J. Khoury and B. A. Ovrut,  Phys. Rev. D {\bf{76}} 123503 (2007)
[hep-th/0702154];
P. Creminelli and L. Senatore,  J. Cosmol. Astropart. Phys. {\bf{0711}} 010 (2007) [hep-th/0702165].



\bibitem{Ell_Mads} G. F. R.
Ellis and M. S. Madsen,   Class. Quantum Grav. {\bf{8}} 667 (1991).


\bibitem{Collinsandellis} C. B. Collins and G. F. R. Ellis, ``Singularities in Bianchi cosmologies", Physics Reports {\bf{56}} 65-105 (1979).


\bibitem{WE}
J. Wainwright and G. F. R. Ellis, ``Dynamical systems in cosmology'' (Cambridge
University Press, Cambridge, 1997).




\bibitem{AnderssonMoncrief}
L. Andersson,
``Cosmological Models and Stability '', in 
 ``General Relativity, Cosmology and Astrophysics, Fundamental Theories of Physics'',  {\bf{177}} p. 277 
(Springer International Publishing Switzerland, 2014; ISBN 978-3-319-06348-5);
L. Andersson and V. Moncrief, ``Future complete vacuum spacetimes'', in  
``The Einstein equations and the large scale behavior of gravitational fields'',
pp. 71--120 (Birkhäuser, Basel,  2004) [gr-qc/0303045]. 


\bibitem{BarrowTipler}
J. D. Barrow, G. J. Galloway and F. J. Tipler, 
Mon. Not. R. Astron. Soc. {\bf{223}}  835 (1986).

\bibitem{LinWald}
X. Lin and R. M. Wald, 
Phys. Rev. D {\bf{40}} 3280 (1989) \& 
{\bf{41}} 2444 (1990). 


\bibitem{Rendall95}  A. D. Rendall, 
Math. Proc. Camb. Phil. Soc. {\bf{118}}  511
(1995).



\bibitem{Jensen} L. G.
Jensen and  J. A. Stein-Schabes,
Phys. Rev.
D {\bf{35}} 1146 (1987).



\bibitem{infra}
A. M. Polyakov, “Infrared instability of the de Sitter
space,” [arXiv:1209.4135];
A. M. Polyakov, Nucl. Phys. B {\bf{797}}, 199 (2008)
[arXiv:0709.2899];
E. Mottola, Phys. Rev. D {\bf{33}}, 1616 (1986);
N. C. Tsamis and R. P. Woodard, Nucl. Phys. B {\bf{474}}, 235 (1996);
L. R. W. Abramo, R. H. Brandenberger and
V. F. Mukhanov, Phys. Rev. D {\bf{56}}, 3248
(1997) [gr-qc/9704037];





          
          
          
          
          
           

\bibitem{EllisandKing} G. F. R.
Ellis and A. R. King, Commun. Math. Phys.  {\bf{38}} 119 (1974).




\bibitem{art:LK63}
E.~M. Lifshitz and I.~M. Khalatnikov,
Adv. Phys. {\bf 12} 185 (1963);
V.~A. Belinskii, I.~M. Khalatnikov, and E.~M. Lifschitz,
Adv. Phys. {\bf 19}, 525 (1970); {\em{ibid.}}
{\bf 31} 639 (1982);
V.~A. Belinskii and I.~M. Khalatnikov, 
Soviet Scientific Review Section A: Physics Reviews {\bf 3} 555 (1981).

\bibitem{Berger}
B. K. Berger and V. Moncrief,  Phys. Rev. D {\bf{48}} 4676 (1993); B. K. Berger, Living Rev. Rel. 
{\bf{5}} 1 (2002).


\bibitem{DavidG}
D. Garfinkle, Phys. Rev. Lett. {\bf{93}} 161101 (2004);
D. Garfinkle,   Class. Quant. Grav. {\bf{24}} S295 (2007).



\bibitem{bianchi}
J. M. Heinzle and C. Uggla,  
Class. Quant. Grav. {\bf{26}}
075016 (2009);
H. Ringstrom,
Class.
Quant. Grav. {\bf{17}} 713 (2000) \&
Annales Henri Poincare {\bf{2}} 405 (2001);
B. Brehm, ``Bianchi VIII and IX vacuum cosmologies: Almost every solution forms particle horizons and converges to the Mixmaster attractor''
[arXiv:1606.08058, 2016].



\bibitem{Uggla03} C. Uggla, H. van Elst, J. Wainwright and G. F. R. Ellis, 
Phys. Rev. D {\bf{68}} 103502 (2003).



\bibitem{Andersson}
L. Andersson {\em{et al.}}, 
Phys. Rev. Lett. {\bf{94 }}051101 (2005).


\bibitem{Goode} S. W.
Goode and  J. Wainwright, 
Class. Quant. Grav. {\bf{2}} 99 (1985);
S. W. Goode, A. A. Coley and J. Wainwright,
Class. Quant. Grav. {\bf{9}} 445 (1992)
[arXiv:0810.3744]

\bibitem{Claudel} C. M.
Claudel and K. P. Newman, 
Proc. R.
Soc. London, Ser. A {\bf{454}} 3 (1998); R. P. A. C.
Newman,
Proc. R. Soc. London, {\bf{443}} A473 \& A493 (1993). 
 
 
\bibitem{Anguige} K.
Anguige, and  K. P. Tod, 
Ann. Phys. (N. Y.) {\bf{276}} 257
(1999).


\bibitem{Middleton}
J. Middleton and J. D. Barrow,
Phys. Rev. D {\bf{77}} 10352 (2008) [arXiv:0801.4090].




\bibitem{Kirnos}
I. V. Kirnos, A. N. Makarenko, S. A. Pavluchenko and A. V. Toporensky,
Gen. Rel. Grav. {\bf{42}} 2633 (2010) [arXiv:gr-qc/0906.0140].

\bibitem{BarrowHervik}
J. D. Barrow and S. Hervik,
Phys. Rev. D {\bf{81}} 023513 (2010) 
[arXiv:0911.3805] 





\bibitem{Freese}
K.~Freese,
``Status of Dark Matter in the Universe'' [arXiv:1701.01840].





\bibitem{XuHuang}
H. Xu, Z. Huang, Z. Liu and H. Miao, ``Flatness without CMB - the 
Entanglement of Spatial Curvature and Dark Energy Equation of State'' 
[arXiv:1812.09100].



\bibitem{PR2018} C.-G.
Park and B. Ratra, ``Measuring the Hubble constant and spatial curvature from supernova apparent magnitude, baryon acoustic oscillation, and Hubble parameter data'' [arXiv:1809.03598].

\bibitem{YuRyan} H.
Yu, B. Ratra and F.-Y. Wang, Ap. J. {\bf{856}} 3 (2018); J. Ryan, S. Doshi and B. Ratra,  Mon. Not. R. Astron. Soc. {\bf{480}} 759 (2018).


 \bibitem{Larena09template} 
  J.~Larena, J.-M.~Alimi, T.~Buchert, M.~Kunz and P.~S.~Corasaniti, Phys. Rev. D {\bf {79}}, 
083011 (2009) [arxiv:0808.1161].


\bibitem{DiDio}
E.~Di~Dio, F.~Montanari, A.~Raccanelli, R.~Durrer, M.~Kamionkowski and
  J.~Lesgourgues,  J. Cosmol. Astropart. Phys.
  {\bf {1606}} 013  (2016) [arxiv:1603.09073];
C.~D. Leonard, P.~Bull and R.~Allison, Phys. Rev D {\bf{94}} 023502  (2016) [arxiv:1604.01410].



\bibitem{Jimenez} R.
Jimenez,  A. Raccanelli,  L. Verde and  S. Matarrese,  J. Cosmol. Astropart. Phys. {\bf{1804}} 002 (2018).



\bibitem{Clarksonprl}
C.~{Clarkson}, B.~{Bassett} and T.~H. {Lu}, 
Phys. Rev. Lett. {\bf 101} 011301 (2008).




\bibitem{Witten2001}
E. Witten, ``The cosmological constant from the
viewpoint of string theory'', in 
Sources and Detection of Dark Matter and Dark Energy
in the Universe, ed. D. B. Cline pages 27--36 (Springer, Berlin, Heidelberg,
2001).


\bibitem{Steinhardt}
P. Steinhardt and N. Turok,   
Science {\bf{312}}  1180  (2006) [arXiv:astro-ph/0605173].



\bibitem{Weinberg1989}
S. Weinberg, 
Rev. Mod. Phys. {\bf{61 }} 1 (1989).


\bibitem{Padilla}
A. Padilla, ``Lectures on the Cosmological Constant
Problem'' [arXiv:1502.05296].


\bibitem{Riess}
A. G. Riess {\em{et al.}},  
Astron. J. {\bf{116}} 1009 (1998).

\bibitem{Perlmutter}
 S. Perlmutter {\em{et al.}}, 
Astrophys.
J. {\bf{517}} 565 (1999).





\bibitem{bubbles}
C. L. Wainwright,  M. C. Johnson,  A. Aguirre and H. V. Peiris,  
 J. Cosmol. Astropart. Phys. {\bf{1410}} 030,  (2014) [arXiv:1407.2950];
C. L. Wainwright, M. C. Johnson, H. V. Peiris, A. Aguirre, and L. Lehner,
J. Cosmol. Astropart. Phys. {\bf{2014(03)}}  (2014) [arXiv:1312.1357].


\bibitem{infl}  W.  E.  East, M.  Kleban, A.  Linde and L.  Senatore, 
J. Cosmol. Astropart. Phys. {\bf{1609}} 010 (2016) [arXiv:1511.05143]; 
J. Braden, M. C. Johnson, H. V. Peiris and A. Aguirre,
Phys. Rev. D {\bf{96}} 023541 (2017) [arXiv:1604.04001];
K. Clough, E. A. Lim, B. S. DiNunno, W. Fischler,
R. Flauger and S. Paban,  J. Cosmol. Astropart. Phys. {\bf{1709}} 025 (2017).




\bibitem{Friedrich1986} H.
Friedrich,
J. Geom. Phys. {\bf{3}} 101 (1986).


\bibitem{Wald83} R. Wald,
 Phys. Rev. D {\bf{28}} 2118 (1983).

\bibitem{Coleybook}
A. A. Coley, 
``Dynamical systems and cosmology'' (Kluwer Academic,
Dordrecht: ISBN 1-4020-1403-1, 2003). 

\bibitem{exppot} J. M.
Heinzle and A. D. Rendall, 
Comm.
Math. Phys. {\bf{269}} 1 (2007); H.
Ringstrom,
Comm. Math. Phys. {\bf{290}} 155 (2009).





\bibitem{art:ColeyLim2012}
A.~A. Coley and W.~C. Lim, 
Phys. Rev. Lett. {\bf 108} 191101 (2012) [arXiv:1205.2142]; W.~C. Lim and A.~A. Coley,
Class. Quant. Grav. {\bf {31}} 015020 (2014) [arXiv:1311.1857].


\bibitem{Av} 
R. van den Hoogen,  J. Math. Phys. {\bf{50}} 082503 (2009).

\bibitem{Coley10} A. A. Coley,
Class. Quant. Grav. {\bf 27} 245017 (2010) [arXiv:0908.4281].



\bibitem{Averaging} R. M. ~Zalaletdinov,
Gen. Rel. Grav. {\bf 24} 1015 (1992) \&
Gen. Rel. Grav. {\bf 25}  673  (1993) [arXiv:gr-qc/9703016];
M. Mars and R. M. Zalaletdinov, J. Math. Phys. {\bf 38}  4741 (1997).


\bibitem{CPZ}
A. A.~Coley, N. ~Pelavas and R. M.~Zalaletdinov,
Phys. Rev. Letts. {\bf 95} 151102  (2005) [arXiv:gr-qc/0504115].



\bibitem{bu00}
T.~Buchert,
Gen. Rel. Grav. {\bf {32}} 105  (2000) [arXiv:gr-qc/9906015] \&
Gen. Rel. Grav. {\bf {33}}  1381 (2001) [arXiv:gr-qc/0102049].




\bibitem{web}
J.~Einasto,
``Yakov Zeldovich and the Cosmic Web Paradigm'',
in Proc. IAU Symp. {\bf 308}, eds.
R.~van de Weygaert, S.~Shandarin, E.~Saar, J.~Einasto 
(Cambridge Univ. Press, 2017)
[arXiv:1410.6932].

\bibitem{HV1}
F.~Hoyle and M. S.~Vogeley,
Astrophys. J. {\bf 566}  641 (2002) [arXiv:astro-ph/0109357];
Astrophys. J. {\bf 607}  751 (2004) [arXiv:astro-ph/0312533].

\bibitem{Pan11}
D. C.~Pan, M. S.~Vogeley, F.~Hoyle, Y. Y.~Choi, and C.~Park,
Mon. Not. R. Astron. Soc. {\bf 421}  926 (2012) [arXiv:1103.4156].


\bibitem{sdb12}
M.~Scrimgeour {\em{et al.}}, 
Mon. Not. R. Astron. Soc.  {\bf 425} 116  (2012) [arXiv:1205.6812].

\bibitem{h05}
D. W.~Hogg, D. J.~Eisenstein, M. R.~Blanton,
N. A.~Bahcall, J.~Brinkmann, J. E.~Gunn and D. P.~Schneider,
Astrophys. J. {\bf 624}  54 (2005) [arXiv:astro-ph/0411197].


\bibitem{NotGW} 
T.  Buchert {\em{et al.}},  Class.  Quant. Grav.  {\bf{32}}  215021 (2015)
[arXiv:1505.07800].



\bibitem{LIGO}[The LIGO Scientific Collaboration, the Virgo Collaboration]
B. P. Abbott {\em{et al.}}, Phys. Rev. Lett. {\bf{116}} 061102 (2016).

\bibitem{LIGO2} [The LIGO Scientific Collaboration, the Virgo Collaboration] B. P. Abbott {\em{et al.}} Phys. Rev. Lett. 
{\bf{116}}  241102 \& 241103 (2016), Phys. Rev.  Lett. {\bf{118}}
221101 (2017) \& {\bf{119}} 141101 (2017); Astrophys. J. {\bf{851}} L35 (2017); Phys. Rev. Lett. {\bf{123}}, 011102 (2019); ``A Gravitational-Wave Transient Catalog of Compact Binary Mergers Observed by LIGO and virgo during the First and Second Observing Runs'' [arXiv:1811.12907].



\bibitem{YangPaschalidis}
H. Yang, V. Paschalidis, K. Yagi, L. Lehner, F. Pretorius and N. Yunes, 
Phys. Rev. D {\bf{97}} 024049 (2018) [arXiv:1707.00207].










\bibitem{Barack} 
L. Barack {\em{et al.}},
``Black holes, gravitational waves and fundamental physics: a roadmap''
[arXiv:1806.05195].





\bibitem{Bolejko18} K. Bolejko, 
Phys. Rev. D {\bf{97}} 103529 (2018) [arXiv:1712.02967].


\bibitem{Bolejko2011a} 
K. Bolejko, 
 J. Cosmol. Astropart. Phys. {\bf{02}}  025 (2011).



\bibitem{Hamber} H. W. Hamber, Quantum Gravitation, Springer Tracts in Modern Physics (Springer 
Publishing, Berlin and New York, 2009) \&  [arXiv:1707.08188]];
H. W. Hamber and L. H. Sunny Yu, ``Gravitational Fluctuations as an Alternative to Inflation'' [arXiv:1807.10704].



\bibitem{Martin}
J. Martin, ''The Theory of Inflation''
[arXiv:1807.11075].  




\bibitem{Brand} R. Brandenberger, L L. Graef, G. Marozzi and G. P. Vacca, 
``Back-Reaction of Super-Hubble Cosmological Perturbations Beyond Perturbation
Theory''
[arXiv:1807.07494];
R. H. Brandenberger, “Backreaction of cosmological perturbations” [hep-th/0004016].





\bibitem{Wei76}
S. Weinberg, Ap. J. {\bf{208}} L1 (1976).

\bibitem{KP}
 N. Kaiser and J. A. Peacock,     Mon. Not. R. Astron. Soc. {\bf{455}} 4518 (2015)
[arxiv:1503.08506].




\bibitem{Durrer} R. Durrer, {\em{The cosmic microwave background}} (Cambridge: Cambridge University Press, 2008). 



  

\bibitem{CCCS}
C.~Clarkson, T.~Clifton, A.~Coley and R.~Sung, Phys. Rev. D
{\bf{85}} 043506  (2012) [arxiv:1111.2214];
B.~Santos, A.~A.~Coley, N.~C.~Devi and J.~S.~Alcaniz,
J. Cosmol. Astropart. Phys. {\bf {1702}} 047 (2017) [arxiv:1611.01885];
A. A. Coley, B. Santos and V. A. A. Sanghai, J. Cosmol. Astropart. Phys. 
{\bf{05}} 039 (2019) [arXiv:1808.07145].

  
 \bibitem{R18} 
  A.~G.~Riess {\em{et al.}}, Astrophys. J.  {\bf {861}}, 126 (2018) [arxiv:1804.10655].
  
   
 \bibitem{VMS1} 
  E.~Di Valentino, A.~Melchiorri and J.~Silk,
 Phys. Lett. B {\bf {761}}, 242 (2016)
 [arxiv:1606.00634];
  E.~Di Valentino, A.~Melchiorri and O.~Mena, Phys. Rev. D {\bf {96}} 043503  (2017) [arxiv:1704.08342];  E.~Di Valentino, E.~V.~Linder and A.~Melchiorri,
Phys. Rev. D {\bf {97}} 043528 (2018) [arxiv:1710.02153];
J.~Sol, A.~Gmez-Valent and J.~de Cruz Prez,
Phys. Lett. B {\bf {774}} 317 (2017) [arxiv:1705.06723].
   

\bibitem{EastWojtak}
W. East, R. Wojtak and T. Abel, Phys. Rev. D {\bf{97}} 043509 (2018);
J. Adamek, M. Gosenca and S. Hotchkiss, Phys. Rev. D {\bf{93}} 023526 (2016).

\bibitem{tracefree}  G. F. R. Ellis,  H. Van Elst, J. Murugan and J. P Uzan,
Class. Quant. Grav. {\bf{28}} 225007 (2011); G. F. R. Ellis, Gen. Rel. Grav. {\bf{46}} 1619 (2014).










\bibitem{Ijjas}
A.  Ijjas, F. Pretorius and P. J. Steinhardt,
``Stability and the Gauge Problem in Non-Perturbative Cosmology''
[arXiv:1809.07010]; see also 
F. Pretorius. Class.
Quant. Grav. {\bf{22}} 425 (2005) \& D. Garfinkle, Phys. Rev.,
D {\bf{65}} 044029 (2002).

\bibitem{KodamaSasaki}
H. Kodama and M. Sasaki, Prog. Theor. Phys.
Suppl. {\bf{78}} 1–166 (1984).



\bibitem{Carneiro}
S. Carneiro, P. C. de Holanda, C. Pigozzo, F. Sobreira, 
``Is the $H_0$   tension suggesting a 4th neutrino's generation?''
 [arXiv:1812.06064].  

\bibitem{CarrColey}
B. J. Carr and A. A. Coley, Int. J. Mod. Phys. D {\bf{20}} 2733 (2011); 
T. Clifton, B. Carr and A. Coley,
Class. Quantum Grav. {\bf{34}} 135005  (2017) [arXiv:1701.05750].


\bibitem{Gauss} J.
Maldacena, J. High En. Phys. 
{\bf{5}} 013 (2003); G.
Cabass, E. Pajer and F. Schmidt,   J. Cosmol. Astropart. Phys. {\bf{1}}
003 (2017).

\bibitem{Amaro} P.
Amaro-Seoane, H. Audley, S. Babak,  {\em{et al.}}, ``Laser Interferometer Space Antenna'' [arXiv:1702.00786].


\bibitem{CarrHawking} B.J.
Carr and S.W. Hawking, Mon. Not. R. Astron. Soc. {\bf{168}} 399 (1974).


\bibitem{DolgovSilk} A.
Dolgov and J. Silk, Phys.
Rev. D {\bf{47}} 4244 (1993).

\bibitem{Caldwell}
R. R. Caldwell and C. Devulder, Phys. Rev. D {\bf{97}} 023532  (2018) [1706.03765];
R. R. Caldwell, T. L. Smith and  D. G. E. Walker, 
``Using a Primordial Gravitational Wave Background to Illuminate New Physics''
[arXiv:1812.07577]

\bibitem{LISA}
H. Audley {\em{et al.}} (LISA) [arxiv:1702.00786].


\bibitem{Celoria}
 M. Celoria and S. Matarrese,  ``Primordial Non-Gaussianity''
[arXiv:1812.08197].

\bibitem{Carr16} B.
Carr, F. Kuhnel and  M. Sandstad,
Phys.
Rev. D {\bf{94}} 083504 (2016).

\bibitem{CarrSilk} B.
Carr and J. Silk, Mon. Not. R. Astron. Soc. {\bf{478}} 3756  (2018).


\bibitem{GilMarin} H.
Gil-Marin {\em et al.}, Mon. Not. R. Astron. Soc. {\bf{465}} 1757 (2017).



\bibitem{MalikWands}
K. Malik and D. Wands, 
Phys. Rept. {\bf{475}} 1 (2009).


\bibitem{Sanghai} V. A. A. 
Sanghai and  T. Clifton, Class. Quant. Grav. {\bf{34}}
065003 (2017); see also V. A. A. Sanghai and T. Clifton,  Phys. Rev. D {\bf{91}} 103532
(2015), Phys. Rev. D {\bf{93}} 089903 (2016) \& Phys. Rev. D {\bf{94}} 023505
(2016).


\bibitem{Goldberg}
S. R. Goldberg, T. Clifton and K. Malik,
Phys. Rev. D {\bf{95}} 043503 (2017);
S. Goldberg, C. Gallagher and T. Clifton, Phys.
Rev. D {\bf{96}} 103508 (2017).


\bibitem{MalikWandssecond} K. Malik and D. Wands, Class. Quant. Grav. {\bf{21}} L65
(2004);
K. Nakamura, Prog. Theor. Phys. {\bf{110}} 723 (2003) {\em{ibid.}} {\bf{113}} 481 (2005) {\em{ibid.}}  {\bf{117}} 17 (2007).



\bibitem{ClarksonUmehMaartens} C.
Clarkson, O. Umeh, R. Maartens and R. Durrer,  J. Cosmol. Astropart. Phys. {\bf{11}} 036 (2014)
[arxiv:1405.7860].




\bibitem{Bonvin15} C.
Bonvin, C. Clarkson, R. Durrer, R. Maartens and O. Umeh,  J. Cosmol. Astropart. Phys. {\bf{2015}}
050 (2015)  [arxiv:1503.07831];
Bonvin, C. Clarkson, R. Durrer, R. Maartens and O. Umeh,  J. Cosmol. Astropart. Phys. {\bf{1507}} 40 (2015)
[arxiv:1504.01676].


\bibitem{Umeh14} O. Umeh, C. Clarkson,  and R. Maartens, 
Class. Quant. Grav. {\bf{31}} 205001  (2014)  [arxiv:1402.1933].

\bibitem{Ben-Dayan13}  I. Ben-Dayan, M. Gasperini, G. Marozzi, F. Nugier and G. Veneziano,  
 J. Cosmol. Astropart. Phys. {\bf{1306}} 002 (2013) [arxiv:1308.4935].


\bibitem{Fleury17} P. Fleury, C. Clarkson  and R. Maartens,  J. Cosmol. Astropart. Phys. {\bf{1703}} 062  (2017) [arxiv:1612.03726].


\bibitem{BenDayan14}  I. Ben-Dayan, R. Durrer, G. Marozzi and  D. J. Schwarz, Phys. Rev. Lett. {\bf{112}} 221301 (2014)
[arxiv:1401.7973] {\em{ibid.}} 
Phys. Rev. Lett. {\bf{110}}  021301
(2013)  [arXiv:1207.1286].


\bibitem{Rampf} C. Rampf, E. Villa, D. Bertacca and M. Bruni,
Phys. Rev. D {\bf{94}}  083515 (2016) [arXiv:1607.05226]; I. Milillo {\em{et al.}}, Phys. Rev. D {\bf{92}} 023519 (2015).


\bibitem{Bartolo}
N. Bartolo, D. Bertacca, M. Bruni, K. Koyama, R. Maartens and S. Matarrese,
Physics of the dark universe {\bf{13}} 30 (2015) [arXiv:1506.00915].



\bibitem{HawkingNature}
S.W. Hawking, Nature {\bf{248}} 30 (1974). 

\bibitem{BernardCarr} B. J. Carr,
``Primordial black holes as dark matter and
generators of cosmic structure''
"Contribution to Proceedings of Simons Conference "Illuminating Dark Matter", held in Kruen, Germany, in
May 2018, eds. R. Essig, K. Zurek, J. Feng (to be published by Springer).


\bibitem{Carr80} B. J. Carr, Astron. Astrophys. {\bf{89}} 6 (1980).
 
\bibitem{Bardeen}
J. M. Bardeen, Phys. Rev. D {\bf{22}} 1882 (1980);
V.F. Mukhanov, H. A. Feldman and R. H. Brandenberger,  Phys. Reports {\bf{215}} 203 (1992).


\bibitem{gaugeinv} S. W. Hawking, Astrophys. J. 
{\bf{145}} 544 (1966);
G. F. R. Ellis and M. Bruni, Phys. Rev. D {\bf{40}} 1804 (1989);
M. Bruni, P. K. S. Dunsby and G. F. R. Ellis, Astrophys. J. {\bf{395}} 34 (1992); 
M. Bruni, S. Matarrese, S. Mollerach and S. Sonego, Class. Quant. Grav. {\bf{14}}
2585 (1997). 

\bibitem{SachsWolfe} 
R. K. Sachs and A. M. Wolfe,  Astrophys. J. {\bf{147}} 73 (1967). 

\bibitem{Ratra2017} B.
Ratra,  Phys. Rev. D {\bf{96}} 103534 (2017);
C.-G. Park and B. Ratra, ``Using the tilted flat and  non-flat inflation models to measure cosmological parameters from a compilation of observational data'' [arXiv:1801.00213];
C.-Z. Ruan, M. Fulvio, C. Yu and Z. Tong-Jie, ``Using spatial curvature with HII galaxies and cosmic chronometers to explore the tension in $H_0$'' 
[arXiv:1901.06626].

\bibitem{Ryan2019}
J. Ryan, Y. Chen and B. Ratra, ``Baryon acoustic oscillation, Hubble parameter, and angular size measurement constraints on the Hubble constant, dark energy dynamics, and spatial curvature''
[arXiv:1902.03196].


\bibitem{Durrer1996}
R. Durrer, Helv. Phys. Acta {\bf{69}} 417 (1996).


\bibitem{Martin19} J. Martin, 
``Cosmic Inflation: Trick or Treat?''
[arXiv:1902.05286]; 
D. Chowdhury, J. Martin, C. Ringeval and V. Vennin, ``Inflation after Planck: Judgment Day'' [arXiv:1902.03951].


\bibitem{GreenWald}
S. R. Green and  R.  M.  Wald,  Class. Quant. Grav. {\bf{31}} 234003 (2014).


\bibitem{infl2}  
M. Kleban and L. gSenatore, 
J. Cosmol. Astropart. Phys. {\bf{10}} 022 (2016)
[arXiv:1602.03520]; A. Linde, 
Found. Phys. {\bf{48}} 1246 (2018) [arXiv:1710.04278].  



\bibitem{siren} 
B. P. Abbott {\em{et al.}}
(LIGO Scientific Collaboration,  the Virgo Collaboration),  Astrophys. J. Lett. {\bf{876}} L7 (2019); C. Guidorzi {\em{et al.}}, Ap. J. Lett.
{\bf{85}} L36 (2017)
[arXiv:1710.06426]. 

\bibitem{Carlip19}
S. Carlip, ``How to Hide a Cosmological Constant''
[arXiv:1905.05216].


\bibitem{generalGWREFS}  C. P. L. Berry {\em{et al.}}, ``The unique potential of extreme mass-ratio inspirals for gravitational-wave astronomy'' [arXiv:1903.03686]; 
S. T. McWilliams {\em{et al.}}, ``Decadal Science White Paper: The state of gravitational-wave astrophysics in 2020'' [arXiv:1903.04592];
 D. Reitze {\em{et al.}}, ``The US Program in Ground-Based Gravitational Wave Science:  Contribution from the LIGO Laboratory, [arXiv:1903.04615]; 
 R. Caldwell {\em{et al.}}, ``Science White Paper: Cosmology with a Space-Based Gravitational Wave Observatory'' [arXiv:1903.04657].








\bibitem{TB}
T. Buchert, P. Mourier and X. Roy, Class. Quant.  Grav. {\bf{35}} 24LT02 (2018)  [arXiv1805.10455];
A. Heinesen, P. Mourier and T. Buchert, ``On the covariance of scalar averaging'' [arXiv:1811.01374]. 



\bibitem{Bolejko2017} 
K. Bolejko, 
 J. Cosmol. Astropart. Phys. {\bf{06}} 025 (2017).

                                                           
\bibitem{VPG}
I. Brown, A. Coley and J. Latta, Phys Rev D. {\bf 87} 043518 (2013)[arXiv:1211.0802];
I. A. Brown, A. A. Coley, D. L. Herman and J. Latta, Phys. Rev. {\bf 88} 083523  (2013) [arXiv:1308.5072]. 

\bibitem{Roukema19}
B. F. Roukema, J. J. Ostrowski, P. Mourier and Q. Vigneron,
``Does spatial flatness forbid the turnaround epoch of collapsing structures?'', Aston Astrophys
[arXiv:1902.09064].

                   

\bibitem{string}
M Green,  J Schwarz and  E Witten, ``Superstring Theory'' (Cambridge: Cambridge
University Press, 1988);  J. Polchinski, ``String Theory'' (Cambridge: Cambridge University Press 2005)



\bibitem{EmparanReall} 
R. Emparan and H. S. Reall,
Living Rev. Rel. {\bf{11}} 6 (2008) [arXiv:0801.3471].




\bibitem{Galloway}
G. J. Galloway and J. M. M. Senovilla, 
Class. Quant. Grav. {\bf{27}}  152002 (2010).


\bibitem{SchoenYau2017}
R. Schoen and S.-T. Yau. ``Positive Scalar Curvature and Minimal Hypersurface Singularities''
[arXiv:1704.05490].



\bibitem{LehnerPretorius2010}
L. Lehner and F. Pretorius,  
Phys. Rev. Lett. {\bf{105}} 101102
(2010).


\bibitem{GregoryLaflamme}
R. Gregory and R. Laflamme,  
Phys. Rev. Lett. {\bf{70}} 2837
(1993); J. E. Santos and B. Way, Phys. Rev. Lett. {\bf{114}}, 221101 (2015);
K. Tanabe, J. High En. Phys.  {\bf{02}} 151 (2016);
P. Figueras, M. Kunesch, and S. Tunyasuvunakool, Phys. Rev.
Lett. {\bf{116}} 071102 (2016).



\bibitem{EmparanMyers}
R. Emparan and R. C. Myers, J. High En. Phys.  {\bf{09}} 025 (2003);
O. J. C. Dias, P. Figueras, R. Monteiro, J. E. Santos, and
R. Emparan, Phys. Rev. D   {\bf{80}} 111701 (2009);
P. Figueras, M. Kunesch, L. Lehner, and S. Tunyasuvunakool, 
Phys. Rev. Letts. {\bf{118}} 151103 (2017).



\bibitem{Henneaux}
M. Henneaux,
Khalatnikov-Lifshitz analysis 
in {\em{Quantum Mechanics of Fundamental Systems:  the  Quest for Beauty and Simplicity - Claudio Bunster 
Festsschrift}} (Berlin, Springer) [arXiv:0806.4670]. 



\bibitem{mald}  J. M. Maldacena,  
Int. J. Theor. Phys. {\bf{38}} 1113 
(1999); J. M. Maldacena, Adv. Theor. Math. Phys. {\bf{2}} 231 (1998).


\bibitem{Klebanov}
I. Klebanov and J. Maldacena,  
Physics Today {\bf{62}} 
28 (2009).



\bibitem{PiotrBizon} P. Bizon, 
Gen. Rel. Grav. {\bf{46}} 1724 (2014) [arXiv:1312.5544].


\bibitem{Bizon} P. Bizon and A. Rostworowski, 
Phys.
Rev. Lett. {\bf{107}} 031102 (2011).


\bibitem{DiasSantos}
O. J. C. Dias, G. T. Horowitz and J. E. Santos, 
Class. Quant. Grav. {\bf{29}} 194002  (2012)  [arXiv:1109.1825];
O. J. C. Dias, and J. E. Santos,
Class. Quant. Grav. {\bf{33}} 23LT01 (2016) \& ``AdS nonlinear instability: breaking 
spherical and axial symmetries'' [arXiv:1705.03065]; 
A. Rostworowski, 
Class. Quant. Grav. {\bf{33}} 23LT01 (2016)] [arXiv:1612.00042];
O. J. C. Dias, G. T. Horowitz, D. Marolf and  J. E. Santos, Class. Quant. Grav. {\bf{29}}  235019  (2012);
S. R. Green, A. Maillard, L. Lehner and S. L. Liebling,  
Phys. Rev. D {\bf{92}} 084001 (2015) [arXiv:1507.08261].





\bibitem{Martinon} G. Martinon,
``The instability of anti-de Sitter space-time''
 [arXiv:1708.05600].




\bibitem{DeWitt}
 B. S. DeWitt, 
 Phys. Rev. {\bf{160}} 1113 (1967).



\bibitem{Horowitz}
G. T. Horowitz, New J. Phys. {\bf{7}} 201 (2005).

  

\bibitem{maldecena} J. Maldacena, 
Adv. Theor. Math. Phys. {\bf{2}} 231 (1998).

\bibitem{EngelhardtHorowitz}
N. Engelhardt and G. T. Horowitz,
Int. J. Mod. Phys. {\bf{D25}}  1643002  (2016) \& 
Phys. Rev. D {\bf{93}} 026005 (2016).

 


\bibitem{Wall2013}
A. C. Wall, 
Class.
Quant. Grav. {\bf{30}} 165003 (2013).





\bibitem{Singh14}
P. Singh,
Bull. Astr. Soc. India {\bf{42}}  121 (2014)
[1509.09182];
I. Agullo and P. Singh, ``Loop Quantum Cosmology: A brief review''
contribution for a volume edited by A. Ashtekar and J. Pullin, to be published in the
World Scientific series 100 Years of General Relativity (World Scientific, Singapore)
[arXiv:1612.01236]; A. Corichi and P. Singh, Phys. Rev. Lett. {\bf{100}} 161302 (2008).



\bibitem{AshtekarandSingh}
A. Ashtekar and P. Singh, 
Class. Quant. Grav. {\bf{ 28}} 213001 (2011); S. Saini and P. Singh, Class. Quant. Grav. {\bf{34}} 235006 
(2017) \& {\bf{35}} 065014 (2018).



\bibitem{Ewing2010}
P.~Singh and E.~Wilson-Ewing, 
Class. Quant. Grav. {\bf 31},  035010 (2014);
A. Corichi and E. Montoy,
Class. Quant. Grav. {\bf{34}} 054001 (2017).
 E. Wilson-Ewing, 
Phys. Rev. D {\bf{82}} 043508 (2010).



\bibitem{Brizuela}
 D. Brizuela, G. A.  Mena Marugán and T. Pawlowski , Class. Quant. Grav. {\bf{27}} 052001 (2010);
 E. Wilson-Ewing, Class. Quant. Grav.
{\bf{35}} 065005 (2018)
[arXiv:1711.10943];
 M. Bojowald and G. M. Paily, Phys. Rev. D {\bf{87}} 044044 (2013).



\bibitem{Tarrio}
P. Tarrio, M. F. Mendez and G. A. M. Marugan, Phys. Rev. D {\bf{88}} 084050 (2013). 




\bibitem{SainiSingh} S. Saini and P. Singh,
``Generic absence of strong singularities and geodesic completeness in modified LQG'' [arXiv:1812.08937]; see also
B. F. Li, P. Singh and A. Wang, Phys. Rev. D {\bf{97}} 084029 (2018) \& {\bf{98}} 066016 (2018);
I. Agullo, Gen. Rel. Grav. {\bf{50}} 91 (2018).



\bibitem{deS}
G. Obied, H. Ooguri, L. Spodyneiko and C. Vafa, ``De Sitter Space and the Swampland" [arXiv:1806.08362];
U. H. Danielsson and T. Van Riet, ``What if string theory
has no de Sitter vacua?" [arXiv:1804.01120];
G. Dvali and C. Gomez,  Annalen Phys. {\bf{528}}, 68 (2016)
[arXiv:1412.8077]; A. Castro, N. Lashkari and A. Maloney,  Phys. Rev. D {\bf{83}} 124027 (2011)
[arXiv:1103.4620].



\bibitem{baumann} D. Baumann and L. McAllister, {\em{Inflation and String Theory}} (Cambridge Monographs on
Mathematical Physics: Cambridge University Press, 2015)
[arXiv:1404.2601].



\bibitem{KKLMMT} S. Kachru, R. Kallosh, A. D. Linde, J. M. Maldacena, L. P. McAllister and S. P. Trivedi,  J. Cosmol. Astropart. Phys. {\bf{0310}}  013 (2003) [hep-th/0308055].


\bibitem{Kall}  R. Kallosh, A. Linde and Y. Yamada,
``Planck 2018 and Brane Inflation Revisited''
[arXiv:1811.01023]; 
Y. Akrami, R. Kallosh, A. Linde and  V. Vardanyan
JCAP {\bf{1806}}  041 (2018);
R. Kallosh, A. Linde and D. Roest, J. High En. Phys.  11 198 (2013)
[arXiv:1311.0472].



\bibitem{double} K. Maeda, S. Mizuno and R. Tozuka, 
``$\alpha$-attractor-type Double Inflation''
[arXiv:1810.06914].




\bibitem{Kiefer} C. Kiefer,  Int. Ser. Monogr. Phys. {\bf{124}} 1 (2004) \& Int. Ser. Monogr. Phys. {\bf{136}} 1 (2007) \& Int. Ser.
Monogr. Phys. {\bf{155}} 1 (2012)]; R. Gambini and J. Pullin, “A first course in LQG quantum gravity,” (p.183 Oxford Univ. Press, UK, 2011); C. Rovelli, “Quantum gravity” (p455 Cambridge Univ. Press, UK, 2004).




\bibitem{Louko} A. Bhardwaj, E. J. Copeland and J. Louko,
``Inflation in LQC''
[arXiv:1812.06841].

\bibitem{Arkani-Hamed}
N. Arkani-Hamed, L. Motl, A. Nicolis and C. Vafa,
J. High En. Phys.  {\bf{0706}} 060 (2007) [hep-th/0601001];
C. Cheung and G. N. Remmen,  Phys. Rev. Lett. {\bf{113}} 051601 (2014)
[arXiv:1402.2287].


\bibitem{Denef}
F. Denef, A. Hebecker and T.Wrase, ``The dS swampland
conjecture and the Higgs potential" [arXiv:1807.06581];
D. Andriot, ``New constraints on classical de Sitter:
flirting with the swampland" [arXiv:1807.09698];
C. Roupec and T. Wrase, ``de Sitter extrema and the
swampland" [arXiv:1807.09538];
A. Kehagias and A. Riotto, ``A note on Inflation and the
Swampland" [arXiv:1807.05445];
J. L. Lehners, ``Small-Field and Scale-Free: Inflation and
Ekpyrosis at their Extremes" [arXiv:1807.05240].





\bibitem{Heisenberg}
L. Heisenberg, M. Bartelmann, R. Brandenberger, A. Refregier, 
Phys. Rev. D {\bf{98}} 123502 (2018) [arXiv:1808.02877];
Y. Akrami, R. Kallosh, A. Linde and V. Vardanyan,
``The landscape, the swampland and the era of precision cosmology''
[arxiv:1808.09440].






\end{thebibliography}
\end{document}